\documentclass[aps,prb,amsmath,twocolumn,amssymb,titlepage,reprint]{revtex4-2}

\usepackage{graphicx}
\usepackage{siunitx}
\usepackage{xcolor}
\usepackage{amsmath}
\usepackage[colorlinks=true]{hyperref}
\usepackage{lipsum}
\usepackage{float}

\definecolor{ccol}{rgb}{0,0.59,0.51}
\definecolor{ucol}{rgb}{.27,.39,.67}
\definecolor{red}{cmyk}{0.25,1,1,0}
\hypersetup{citecolor=ccol, urlcolor=ucol}

\begin{document}
    \title{Investigation of Purcell enhancement of quantum dots emitting in the telecom O-band with an open fiber-cavity}
    
    \author{J. Maisch$^{1,2}$}
    \thanks{These two authors contributed equally}
    \author{J. Grammel$^{2,3}$}
    \thanks{These two authors contributed equally}
    \author{N. Tran$^{1,2}$}
    \author{M. Jetter$^{1,2}$}
    \author{S.~L. Portalupi$^{1,2}$}
    \author{D. Hunger$^{2,3,4}$}
    \author{P. Michler$^{1,2}$}
    \email{p.michler@ihfg.uni-stuttgart.de}
    \affiliation{$^1$ Institut für Halbleiteroptik und Funktionelle Grenzflächen, SCoPE, University of Stuttgart, Allmandring 3, 70569 Stuttgart, Germany}
    \affiliation{$^2$ Center for Integrated Quantum
    	Science and Technology (IQ\textsuperscript{ST}), University of Stuttgart, Allmandring 3, 70569 Stuttgart, Germany}
    \affiliation{$^3$ Physikalisches Institut (PHI), Karlsruhe Institute of Technology (KIT), Wolfgang-Gaede Str. 1,
        Karlsruhe 76131, Germany}
    \affiliation{$^4$ Institute for Quantum Materials and Technologies (IQMT), Karlsruhe Institute of Technology (KIT), Herrmann-von-Helmholtz Platz 1, 76344 Eggenstein-Leopoldshafen}

    \begin{abstract}        
        Single-photon emitters integrated in optical micro-cavities are key elements in quantum communication applications. However, optimizing their emission properties and achieving efficient cavity coupling remain significant challenges. In this study, we investigate semiconductor quantum dots (QDs) emitting in the telecom O-band and integrate them in an open fiber-cavity. Such cavities offer spatial and spectral tunability and intrinsic fiber-coupling. The design promises high collection efficiency and enables the investigation of multiple emitters in heterogeneous samples. We observe a reduction of the decay times of several individual emitters by up to a factor of $\num{2.46(2)}$ due to the Purcell effect. We comprehensively analyze the current limitations of the system, including cavity and emitter properties, the impact of the observed regime where cavity and emitter linewidths are comparable, as well as the mechanical fluctuations of the cavity length. The results elucidate the path towards efficient telecom quantum light sources.
    \end{abstract}

    \maketitle

    %
    %
    %
    %
    %
    \section{Introduction}
The technological realization of quantum communication networks relies on versatile single-photon sources as a key building block. Recent research has explored various approaches to these quantum emitters \cite{Gisin2007,Vajner2022}. In the context of technological applications, there is particular interest in the emission in the telecom wavelength regime, which facilitates long-distance transmission through optical fibers. Furthermore, emitters embedded into optical micro-cavities are intensively studied and were demonstrated to show high emission performances in terms of brightness \cite{Tomm2021,Nawrath2023}, single-photon purity \cite{Hanschke2018} and indistinguishability \cite{Santori2002,Somaschi2016,Weiler2010,Tomm2021}. Within micro-cavities, the emission is directed into the cavity mode and enhanced due to the Purcell effect \cite{Purcell1946}.

    One promising platform for single-photon emission are semiconductor quantum dots (QDs) \cite{Michler2017}. Especially the
    GaAs/InAs material system has exhibited high-quality emitters in the near-infrared (NIR) spectral regime \cite{Somaschi2016}. In recent years, these emitters were also successfully realized in the telecom
    regime, specifically in the O-band\cite{Ward2005,Paul2015,Dusanowski2017} and the C-band \cite{Paul2017}.
    Furthermore, QDs on the same platform were already successfully integrated into various micro-cavities, such as
    micro-pillars \cite{Somaschi2016}, waveguides \cite{Javadi2015}, circular Bragg gratings \cite{Liu2019,Wang2019,Kolatschek2019}, photonic crystals
    \cite{Phillips2024} or open cavities \cite{Herzog2018,Tomm2021}.

    Among these, open cavities stand out due to their spatial and spectral tunability, enabling the investigation of several different emitters in the same cavity. They also allow for the distinction of cavity and emitter properties, in addition to acquiring comparable data sets to draw conclusions about the reproducibility and the statistical distribution of the fabricated emitters.
    In contrast, most of the other micro-cavity approaches rely on integrated photonic structures with fixed geometries and dedicated cavities for distinct emitters. Therefore, open cavities give manifold research possibilities \cite{Pfeifer2022} in the field of scanning-cavity microscopy
    \cite{Mader2015}, but also allow a transition to integrated, fiber-pigtailed devices \cite{Schlehahn2018, Rickert19,
        Rickert21}. For semiconductor QDs, one of the outstanding emitter performances overall was demonstrated with
    QDs within an open cavity emitting in the NIR regime \cite{Tomm2021, Ding2023}.
    From a technological standpoint, an open cavity based on an optical fiber has the advantage that the emitted light is intrinsically fiber-coupled.
    While for other QD-cavity approaches the transfer to the telecom regime was demonstrated recently \cite{Kolatschek2021, Nawrath2023}, for fiber-cavities, however, this step was still missing.

    In this work, we comprehensively study the emission from QDs in the telecom O-band (around $\lambda_0=\SI{1310}{\nano\metre}$) embedded in an open fiber cavity. By combining intrinsic fiber coupling and the cavity enhancement, we address two key elements relevant for future quantum network applications. We present a systematic characterization of the fundamental cavity and emitter parameters. Additionally, through deterministic preselection of QDs, we directly compare emitter properties inside and outside the cavity for several emitters. We measured the Purcell enhancement, observing decay time reductions of up to a factor of $\num{2.46(2)}$. This value reflects the influences of the emitter and cavity linewidth, the sample quality and the vibrational noise. To model our experimental conditions, we include the limiting effects of finite emitter and cavity linewidths and extend the existing theory of the Purcell enhancement of a fluctuating cavity, achieving good agreement with our measurement results.

    \section{Methods}
\begin{figure}
        \includegraphics[width=\columnwidth]{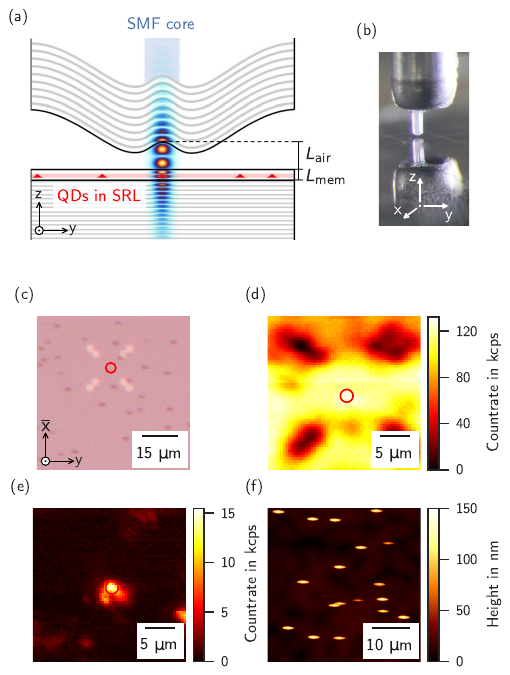}
        \caption{(a) Schematic of the cavity setup: A QD sample with a bottom distributed Bragg reflector (DBR) together with a laser-machined single mode fiber (SMF) with a concave profile and DBR coating form a tunable microcavity. The fiber's cladding is cropped to minimize the impact of small tilts on the minimal cavity length. A simulation of the fundamental cavity mode's energy distribution is shown. (b) Photograph of the setup taken during alignment at room temperature. The fiber is held above the semiconductor sample and is mirrored in the surface. (c) Image of the sample surface captured with an optical microscope. The deterministically fabricated gold markers are visible and the respective QD location is indicated (red circle). (d) Photoluminescence scan (collected through the fiber) of the sample surface to find the QD position between the gold markers. (e) Resonant scanning cavity microscopy image over the same area as in d), spectrally filtered around the QD emission. (f) Atomic force microscopy scan of the sample topography. (c)-(f) are displayed in the same orientation.}
        \label{fig1}
    \end{figure}

     A fiber based Fabry-Pérot cavity was formed by the tip of a optical single mode fiber and a semiconductor sample (see Fig.~\ref{fig1}~(a) and (b)) mounted in a bath cryostat (AttoLiquid). The end facet of the fiber was processed by CO\textsubscript{2}-laser machining \cite{Hunger2010}. The resulting
    concave profile was coated (Laseroptik GmbH, Garbsen, Germany) with a dielectric distributed Bragg reflector (DBR). We used fibers with two different types of coatings corresponding to transmissions of ${T_\text{fiber,1}=\SI{1000}{ppm}}$ and $T_\text{fiber,2}=\SI{100}{ppm}$ (DBR consists of 13 and 17 pairs of Nb\textsubscript{2}O\textsubscript{5}/SiO\textsubscript{2}). The respective radii of curvature were  $RC_\text{fiber,1} = \SI{34.3}{\micro\meter}$ and $RC_\text{fiber,2} = \SI{43.7}{\micro\meter}$. The semiconductor sample was grown by metal-organic
    vapor phase epitaxy (MOVPE). On a GaAs substrate (n-doped), a DBR was grown (35 pairs of AlAs/GaAs,
    $T_\text{DBR}\approx\SI{1500}{ppm}$). After another \SI{193.7}{\nano\metre} ($\approx1/2\cdot\lambda_0/n$) thick GaAs layer, InGaAs QDs were grown in Stranski-Krastanow growth mode. The QDs were topped by an InGaAs layer serving as strain reducing layer (SRL) and
    ensuring the emission in the O-band at around \SI{1310}{\nano\metre} \cite{Paul2015}. On top, a GaAs capping layer
    ($\SI{295.6}{\nano\metre} \approx 3/4\cdot\lambda_0/n$) finished the sample. All layers above the DBR form a thin membrane inside the resonator with the total thickness $L_\text{mem} \approx 5/4\cdot\lambda_0/n$. This membrane thickness fixes the cavity mode to an air-like mode \cite{Janitz2015, Dam2018}. The ratio of the transmissions of the fiber and sample DBR was designed for a QD sample transmission of \SI{100}{ppm} (for 35 pairs of AlAs/GaAs). However, the measured transmission of the sample DBR deviated significantly from the design value. This led to a strongly over-coupled cavity and the designed, optimized outcoupling (for the \SI{1000}{ppm} fiber coating) and the optimized outcoupling with Purcell enhancement (for the \SI{100}{ppm} coating) was not achieved in this experiment.
    
    The QD sample was investigated within two setup configurations: The micro-photoluminescence (\textmu PL) and the scanning cavity microscope. Both were realized in the same bath cryostat. For the cavity setup, the in- and outcoupling of the light is  realized via the utilized fiber. Between the two mirrors the cavity mode forms (Fig.~\ref{fig1}~(a) shows a schematic of the micro-cavity together with a numerical simulation of the cavity mode field). The sample position could be controlled in all three spatial dimensions (x, y, z) by a precise piezo electric positioning system with sub-nanometric resolution (attocube ANS100 series). The same setup was also used for the sample investigation under \textmu PL, where the top fiber is substituted with a microscope objective ($\text{NA}=\num{0.82}$).
    On the grown sample, positions of several suitable QDs were first pre-selected by low temperature in-situ photolithography \cite{Sartison2017}.  
    Subsequently, gold markers were deposited. Figure~\ref{fig1}~(c) shows an image of the sample surface with markers captured by a optical microscope.
    This allows a one-to-one comparison of the same QD under \textmu PL or when placed in the fiber-cavity \cite{Herzog2018}. The use of gold markers blocks the light emission locally making the markers visible when acquiring spatial photoluminescence scans as in Fig.~\ref{fig1}~(d).
    Figure ~\ref{fig1}~(e) shows a resonant scanning cavity microscopy image at a fixed cavity length matching the QD transition. The signal was additionally filtered spectrally using a \SI{1250}{\nano\meter} long-pass to eliminate the background luminescence outside the cavity's stopband. 
    In total, three QDs were studied on this sample in detail, which we name QD A, B and C. If not stated otherwise, all presented measurements were performed within the bath cryostat at \SI{4}{\kelvin}.
    
    Figure~\ref{fig1}~(f) shows an atomic force microscopy (AFM) scan ($\SI{40}{\micro\meter}\times\SI{40}{\micro\meter}$) of an exemplary part of the sample surface. Prominently, there are bumps with a typical height of $\SI{150}{\nano\meter}$ which are also visible in the image captured by the light microscope (Fig.~\ref{fig1}~(c)). Besides the bumps, the rms surface roughness amounts to $S_q=\num{2}\pm\SI{1}{\nano\meter}$ within an area of $\SI{3}{\micro\meter}\times\SI{3}{\micro\meter}$ (larger than the lateral area of the cavity mode).

    \section{Results and Discussion}
    \begin{figure*}
        \includegraphics[width=\textwidth]{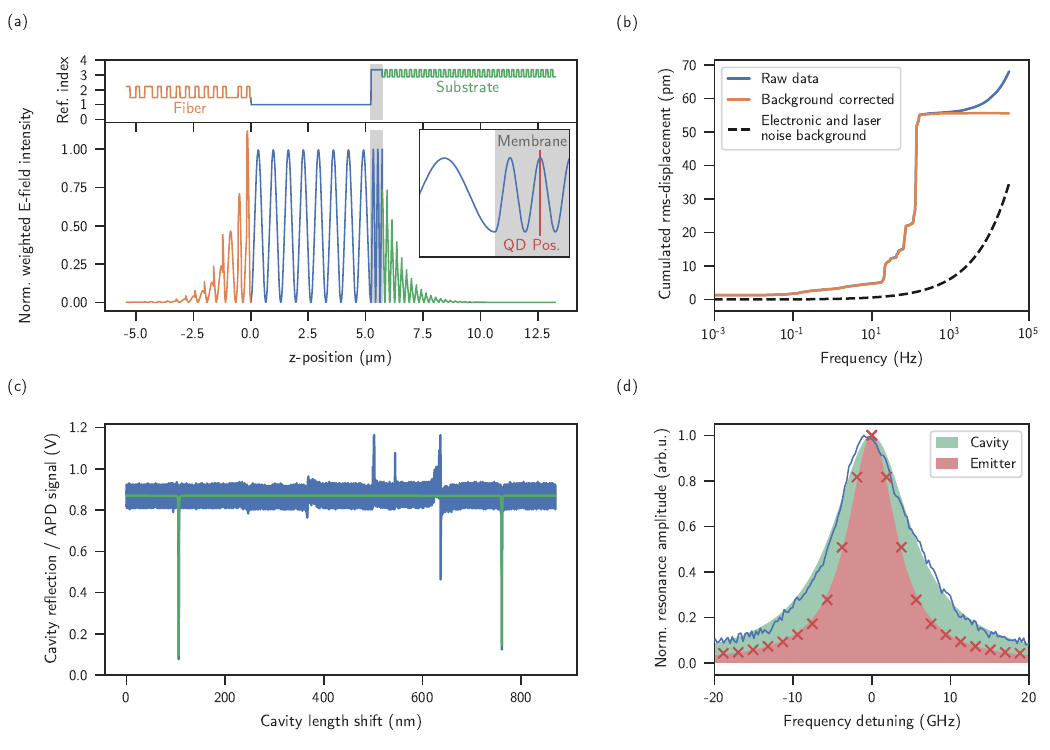}
        \caption{(a) Transfer matrix simulation of the cavity system. The simulation is based on the refractive index profile displayed in the top panel and results in the energy distribution shown in the bottom panel. The chosen dimensions correspond to the longitudinal mode order $m=\num{8}$ in the air gap of the cavity, which is the last mode observable in the experiment when the fiber gets in contact with the substrate. The inset shows the weighted E-field intensity distribution in the membrane (grey area) in detail and the position of the QD is indicated (red vertical line). (b) Cumulated rms displacement as a result of the Fourier analysis calculated for noise estimation. The measured noise baseline (dashed black line), the raw data (blue line) and the background corrected data (orange line) are displayed. (c) Result of a cavity scan over one free spectral range. The two fundamental resonance dips are fitted with Lorentzian fits (green) and a finesse of $\mathcal{F}=\num{1695}$ is extracted. (d) Comparison of the cavity and emitter linewidth. The cavity spectrum (blue line) is obtained by zooming into c) whereas the QD spectrum (red crosses) is a result of a $g^{(1)}$-measurement conducted with a Michelson interferometer. Both line shapes are fitted by Lorentzian fits (FWHM: $\Delta\nu_\text{cav}=\SI{12.2}{\giga\hertz}$ and $\Delta\nu_\text{em}=\SI{7.4}{\giga\hertz}$) displayed by the filled areas (green and red).}
        \label{fig2}
    \end{figure*}
    In a first step, we performed a thorough study of the cavity parameters, both experimentally and by numerical simulations at $\lambda_0 = \SI{1310}{\nano\meter}$. Figure~\ref{fig2}~(a) displays the electric field distribution in the cavity obtained from a simulation with a transfer-matrix model. The top panel shows the
    refractive index profile along the z-axis. All values are adapted from the design parameters of the structures and were used as input for the calculations. On
    the left side the fiber mirror (Nb\textsubscript{2}O\textsubscript{5}/SiO\textsubscript{2}) is represented whereas the right part comprises the QD layer and the semiconductor DBR (AlAs/GaAs). The uneven distribution within the fiber DBR stack leads by design to two stopbands, one around \SI{1310}{\nano\meter} and the second around \SI{1550}{\nano\meter}. In this work, for the O-band quantum dots, only the first stopband is used. Both mirrors are separated by a small air gap (refractive index $n=1$). The bottom panel shows the
    simulated effective energy distribution through the very same structure. By integration, one can extract the effective energy distribution length \cite{Greuter2014, Dam2018}
    \begin{equation}
    	L_\text{eff}= 2\int\frac{|E(z)|^2\, n^2(z)}{|E_\text{GaAs}|^2\, n_\text{GaAs}^2} \,\text{d}z \quad .
    \end{equation}
 	This integral equals the longitudinal part of the integration of the mode volume and the factor of $2$ compensates the filling factor of $1/2$ of the standing wave in longitudinal direction. In Fig.~\ref{fig2}~(a) the 8\textsuperscript{th} longitudinal mode order within the air gap is shown, corresponding to the lowest mode order observable in the experiment when the fiber comes into contact with the substrate. At this mode, an effective energy distribution length of $L_\text{eff,sim}= \SI{7.25}{\micro\meter}$ is simulated. In the experiment, we extract the cavity length at the same resonance as ${L_\text{opt,exp}\approx \SI{9.86}{\micro\meter}}$. Here, the optical cavity length
    \begin{equation}
    	\begin{split}
			L_\text{opt} &= L_\text{pen,fib} + \int_{0}^{L_\text{air} + L_\text{mem}} n(z) \,\text{d}z + L_\text{pen,sc}\\
			&= L_\text{pen,fib} + L_\text{air} + L_\text{mem}\cdot n_\text{GaAs} + L_\text{pen,sc}
    	\end{split}
    \end{equation}
    comprises the optical lengths of the air gap and the membrane (compare Fig.~\ref{fig1}~(a)) as well as the respective frequency penetration depths $L_\text{pen}=c\tau/2$ of the fiber and the semiconductor mirror, where $\tau$ is the group delay \cite{Koks2021}. For the given situation of the 8\textsuperscript{th} resonance we find $L_\text{pen,fib} =\SI{1.05}{\micro\meter}$, $L_\text{air} ={8\cdot\lambda_0/2}=\SI{5.24}{\micro\meter}$, ${L_\text{mem}\cdot n_\text{GaAs}} ={2.5\cdot\lambda_0/2}= \SI{1.64}{\micro\meter}$ and $L_\text{pen,sc} =\SI{1.96}{\micro\meter}$  which sums up to $L_\text{opt,sim} =\SI{9.89}{\micro\meter}$ matching the experimental result to a high degree.

    The simulation of the field distribution shows the optimized design of the DBRs. In particular, the reflection phases of the DBRs are tuned, such that the membrane's interface inside the cavity lies in a field node, minimizing the scattering losses from the growth-related rough semiconductor surface.
    This optimization likewise fixes the cavity  to an air-like mode, with a homogeneous energy distribution over the entire cavity (air gap and membrane) \citep{Dam2018}.
    An alternative approach would have been to position an antinode on the membrane surface, theoretically leading to a higher energy density within the membrane.
    However, the same membrane thicknesses also result in increased transmission through the underlying DBRs, which in turn lowers the finesse. Consequently, we made the deliberate choice to avoid this configuration and the associated additional scattering losses.
    The QD is positioned in a field maximum to maximize the coupling strength (compare inset Fig.~\ref{fig2}~(a)).
    
   In practice, there are mechanical vibrations which induce fluctuations in the cavity resonance frequency (due to cavity length shifts). For one, there is a contribution due to the finite
    stability of the used positioning systems. Additionally, external sources could further increase the noise level. By suitable isolation of the experiment, the external sources can be reduced to a minimum.
    In order to quantify the fluctuations for our system, we measured the reflectance signal of the cavity tuned to the flank of a resonance at $\lambda = \SI{1310}{\nano\meter}$.
    Here, small fluctuations translate into intensity signal variations almost linearly.
    The cumulated noise spectrum shown in Fig.~\ref{fig2}~(b) was obtained via Fourier analysis.
    It can be seen that the major noise contributions are within the range of \num{10}~to~\SI{200}{\hertz}. Most probably, these correspond to mechanical resonances of the positioning system at cryogenic temperatures. Simultaneously, it can be extracted that the total rms-displacement adds up to $\sigma_\text{min}\approx\SI{56}{\pico\meter}$. This value corresponds to the most stable conditions observed in our experiments. In contrast, the maximal jittering of the cavity length was found to be on the order of two cavity linewidths $\sigma_\text{max}\approx\SI{850}{\pico\meter}$.
    Under typical, optimized conditions the jitter is below~$\SI{300}{\pico\meter}$. Recently, stabilities better than $\SI{20}{\pico\meter}$ rms were reached \cite{Fontana2021, Pallmann2023, Fisicaro2024}, although the values crucially depend on the design of the cryo system.
    
    Furthermore, we tuned the cavity length while recording the cavity reflection signal. The cavity length was continuously varied by applying a triangular voltage signal to the z-piezo. The voltage amplitude was chosen such that the scanning range
    covered more than one free-spectral range (FSR) of the cavity. One exemplary result is shown in Fig.~\ref{fig2}~(c). Two
    main dips are visible. These correspond to two consecutive longitudinal cavity resonances which occur whenever the 
    length of the air gap matches the resonance condition given by the incident wavelength ($L_\text{air}=m\lambda/2$, where $m$ denotes the mode number). In between these
    main resonances, other deviations from the signals baseline can be observed. These can be attributed to higher order transverse modes with imperfect
    mode matching~\cite{Gallego2016}. Additionally, a high contrast (up to $C_\text{imp}=\SI{91}{\percent}$) was observed. This indicates excellent impedance matching for fiber 1 (see also Appendix \ref{sec:impedance-matching}).

    By fitting the main resonances in Fig.~\ref{fig2}~(c) with two Lorentzians (shown in green) the finesse can be
    extracted as the fraction of the FSR over the linewidth. Here, $\mathcal{F}=\num{1695}\pm\num{12}$ was found for fiber 1 ($T_\text{fiber,1}=\SI{
    	1000}{ppm}$). In average over several measurements with this fiber, we found a mean finesse of $\num{1788}\pm\num{179}$. For the low-transmittive fiber 2 ($T_\text{fiber,2}=\SI{
    	100}{ppm}$), we found a mean finesse of $\num{3062}\pm\num{47}$.
    	Generally, the finesse is determined by the losses $\mathcal{L}_\text{tot}$ present
    in the cavity on either the semiconductor side (superscript sc) or the fiber side (superscript fib). The finesse can be calculated as \cite[p.~441]{Saleh1991}
    
    \begin{equation}\label{eq:Fintheo}
        \mathcal{F}_\text{theo}= \frac{\pi((1-\mathcal{L}_\text{tot}^\text{sc})(1-\mathcal{L}_\text{tot}^\text{fib}))^\frac{1}{4
        }}{1-((1-\mathcal{L}_\text{tot}^\text{sc})(1-\mathcal{L}_\text{tot}^\text{fib}))^\frac{1}{2}} \quad .
    \end{equation}
	To understand the limitations of our system, we analyzed the different contributions: The transmission losses at the
    mirrors~$\mathcal{L}_\text{trans}$, the scattering losses~$\mathcal{L}_\text{scat}$
    and the losses due to absorption within the material~$\mathcal{L}_\text{abs}$. All parts sum up to the total loss
    \begin{equation}
        \mathcal{L}_\text{tot} = \mathcal{L}_\text{trans} + \mathcal{L}_\text{scat} + \mathcal{L}_\text{abs} \quad .
    \end{equation}
    The losses are determined by the quality of the two cavity mirrors. In a reference setup at room temperature, we combined the fiber tip with a planar dielectric mirror.    
    Using a symmetric cavity with mirror coatings of $T_\text{fiber,2}=\SI{100}{ppm}$ on both sides, we identified the losses associated with the fiber coatings $\mathcal{L}_\text{tot}^\text{fib}$.
    By measuring a reference finesse of \num{27460} (corresponding to total losses of $\SI{228}{ppm} = 2\mathcal{L}_\text{tot}^\text{fib}$ via Eq.~\eqref{eq:Fintheo}), we found that losses other than the transmission are nearly negligible for the fiber coatings.    
    On the other hand, we measured the transmission of the semiconductor sample (QD membrane + DBR) as $\mathcal{L}_\text{trans}^{\text{sc}} = \num{1496}\pm\SI{38}{ppm}$. Additionally, the semiconductor surface introduces significant scattering losses. Via the AFM measurements (see Fig.~\ref{fig1}~(f)) we found an rms-roughness of $S_q=\num{2}\pm\SI{1}{\nano\meter}$. Using this value we estimate the scattering losses \cite{Bennett1961}
    \begin{equation}
        \mathcal{L}_\text{scat}=\left(4\pi\frac{S_q}{\lambda} \right)^2 \quad .
    \end{equation}
    For our target wavelength $\lambda_0 = \SI{1310}{\nano\meter}$, we estimated the maximum scattering losses to be
    $\mathcal{L}_\text{scat}^\text{sc}=\num{368}\pm\SI{92}{ppm}$. The roughness is intrinsically limited due to
    the advanced MOVPE growth in order to obtain QDs emitting in the O-band. In total, we find for coatings $T_\text{fiber,1}=\SI{
    	1000}{ppm}$ and $T_\text{fiber,2}=\SI{
    	100}{ppm}$ theoretical finesse of $\mathcal{F}_\text{theo,1} = \num{2194}$ and $\mathcal{F}_\text{theo,2} = \num{3200}$. This is in good accordance with the typically measured finesses stated above.
    
    In a second step, we studied a quantum emitter within the cavity. The interaction is decisively determined by the linewidth of the emitter and the cavity. In Fig.~\ref{fig2}~(d) one cavity resonance from Fig.~\ref{fig2}~(c) is shown in detail. The x-axis is translated into frequency space and the dip is flipped for comparability with the emitter spectrum. The Lorentzian fit (green area) yields a cavity linewidth as full width half maximum (FWHM) $\Delta\nu_\text{cav} = \num{12.20}\pm\SI{0.09}{\giga\hertz}$. In comparison, also an exemplary spectrum of the emission from QD A is shown in Fig.~\ref{fig2}~(d). The lineshape was obtained via a $g^{(1)}$-measurement with a Michelson interferometer. There a QD linewidth (FWHM) $\Delta\nu_\text{em} = \num{7.4}\pm\SI{0.1}{\giga\hertz}$ was observed (red area).
    
    The observed emitter linewidth is about as large as the cavity linewidth. Therefore, the system is in the transition from the so-called bad cavity regime (${\Delta\nu_\text{cav}\gg \Delta\nu_\text{em}}$) to the bad emitter regime (${\Delta\nu_\text{cav}\ll \Delta\nu_\text{em}}$). In the latter, the cavity acts as a spectral filter for the emitter fluorescence. As a consequence, the indistinguishability of the emitted photons is increased \cite{Grange2015}, whereas the Purcell enhancement is limited to the overlapping fraction of the spectrum which will be discussed below.
    
     \begin{figure}
    	\includegraphics[width=\linewidth]{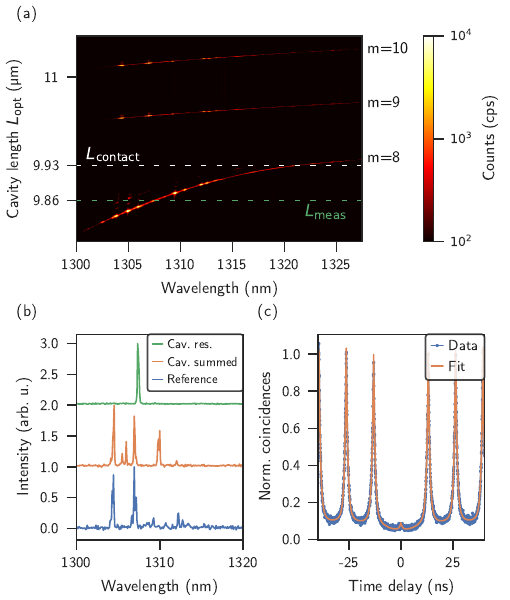}
    	\caption{(a) Dispersion plot as a result of a cavity length scan. $m$~indicates the longitudinal mode order in the air gap. (b) Comparison of spectra: reference (blue), reconstructed within the cavity by summing up all cavity lengths (orange, with offset $+\num{1.0}$)  and one emission line filtered by the cavity at cavity length $L_\text{meas}$ (green, with offset $+\num{2.0}$). (c) $g^{(2)}$-measurement recorded in a Hanbury-Brown and Twiss type setup under pulsed above-band excitation ($f_\text{exc}=\SI{76}{\mega\hertz}$, $\lambda_\text{exc}=\SI{780}{\nano\meter}$)}
    	\label{fig3}
    \end{figure}
\begin{figure}
        \includegraphics[width=\linewidth]{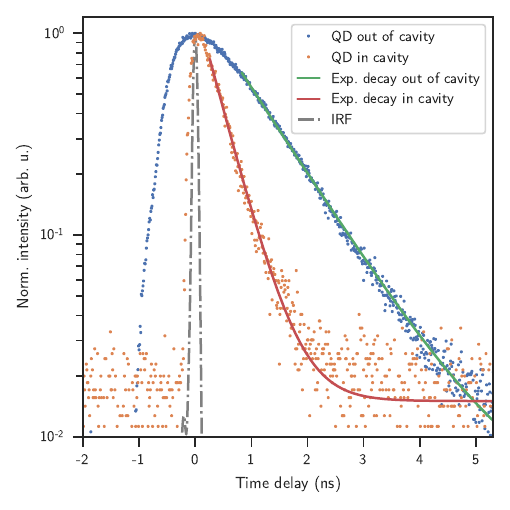}
        \caption{Result of a decay time measurement. Data is shown for both the reference setup (blue dots) and the cavity setup (orange dots). The measurements were performed with the same emission line from QD A. The exponential decays are fitted (solid lines). Additionally, the instrumental response function (IRF) which was considered in the fit is shown (dash-dotted gray line).}
        \label{fig4}
    \end{figure}

    Based on the characterization of the emission spectra, the finesse measurement and the noise analysis, we studied preselected QDs in the cavity. First, the dispersion curve shown in Fig.~\ref{fig3}~(a) was recorded.  A red laser ($\lambda=\SI{615}{\nano\meter}$) was used for above-band excitation of the sample. The emitted signal was recorded with a spectrometer over a spectral range between \SI{1300}{\nano\meter} and \SI{1327}{\nano\meter} (x-axis in Fig.~\ref{fig3}~(a)).
    While lateral position, excitation power and environmental conditions were not changed, the cavity air gap $L_\text{air}$ was varied with a stepsize of \SI{1}{\nano\meter} over a range of about
    \SI{1.5}{\micro\meter} (indicated by the positioning system, proportional to y-axis in Fig.~\ref{fig3}~(a)). Each horizontal line in the plot corresponds to one recorded spectrum at a certain
    cavity length. The color code indicates the respective detected photon counts per second collected from the cavity.
    The chosen z-range contains three cavity modes corresponding to the simulated mode orders in the air gap $m=\SIrange{8}{10}{}$ (indicated in Fig.~\ref{fig3}~(a)). The upper two show an approximately linear evolution within the recorded spectral
    range. This allows to extract the optical cavity length as $L_\text{opt} = \lambda_1\lambda_2/\left(2 \left(\lambda_2 - \lambda_1\right)\right)$ \cite{Hood2001}, where $\lambda_{1/2}$ are two neighboring resonance wavelengths for two consecutive modes in one spectrum in the linear regime. Consequently, a rescaling of the y-axis is possible. The set value of the z-position is translated into a spectrally measured optical cavity length $L_\text{opt}$.
    The lowest mode ($m=\num{8}$), however, clearly differs from a linear slope. This is a clear indication that the fiber
    tip started to touch the semiconductor surface. In that case, the change of the piezo positioner (set value) did no longer
    linearly transfer to the actual optical cavity length. The corresponding length of contact $L_\text{contact} = \SI{9.93}{\micro\meter}$ is indicated in Fig.~\ref{fig3}~(a). Further, a compensation for the nonlinearity of the dispersion allows us to calculate the length after contact. The indicated $L_\text{meas} = \SI{9.86}{\micro\meter}$ corresponds to the conditions of the decay-time measurements presented below (see Fig.~\ref{fig4}).

    Additionally, in Fig.~\ref{fig3}~(a), one can observe high count rates for specific cavity lengths. The bright spots correspond to lengths
    when the cavity is in resonance with a transition from the selected QD. 
    Due to off-resonant cavity feeding,
    there is also emission into the mode besides these particular spots, however at a much lower rate (see the logarithmic color scale). Furthermore, when the fiber gets in contact, the intensity increases since the cavity stability is strongly improved.
	
	Figure~\ref{fig3}~(b) shows a comparison of the spectra of QD~A inside the cavity (orange) and the reference spectrum (blue) acquired in the \textmu-PL configuration.  From the cavity scan, the expected free-space QD spectrum can be reproduced by summing up all individual spectra of the scan. Additionally, a single spectrum (green) from the scan at $L_\text{meas}$ is displayed. To identify the electronic configuration of the present transitions, power and polarization dependent spectra were taken. We expect the QD to emit predominantly from charged states, since a fine structure splitting was not observed. The reference spectrum shows two dominant lines (\SI{1304.4}{\nano\meter} and \SI{1306.9}{\nano\meter}) which both are also clearly visible within the reconstructed cavity spectrum. Inside the cavity, further transitions are increased in intensity probably due to a different polarization of the excitation light. The resonance in the spectrum for fixed length is set to match the selected emission line at $\lambda = \SI{1307.4}{\nano\meter}$. Consequently, only this one line appears in the spectrum. This fact emphasizes the cavity's capability to strongly filter the emission. Notably, when in contact for $m=\num{8}$, a small spectral shift can be observed for all resonances (for example from \SI{1306.9}{\nano\meter} to \SI{1307.4}{\nano\meter}; see also Appendix \ref{sec:strainshift}). Most probably, this can be attributed to increased strain within the semiconductor due to the touching fiber tip.
	
	In figure~\ref{fig3}~(c) a $g^{(2)}$-measurement under pulsed excitation ($f_\text{exc}=\SI{76}{\mega\hertz}$, $\lambda_\text{exc}=\SI{780}{\nano\meter}$) of the investigated QD transition recorded in the cavity is shown. By summing up the coincidences for the central peak and 10 side peaks (normalized to the Poisson level), a raw value of $g_\text{raw}^{(2)}(0)=\num{0.31}$ is obtained. By fitting, $g_\text{fit}^{(2)}(0)=\num{0.04}$ can be extracted which considers the background due to detector dark counts and the peak overlaps. This confirms a high single-photon purity of the photons emitted from the cavity. Additionally, a small bunching due to blinking can be observed on longer time scales (considered for normalization but not shown in the plot). From that, we determine that the QD is in an optically active state \SI{95}{\percent} of time which is a typical value for all investigated transitions within this study.\\ 
		
    A central figure of merit for quantum emitters within a micro-cavity is the Purcell enhancement. One way to observe this is by comparing the decay time of a
    transition inside ($\tau_\text{cav}$) and outside ($\tau_\text{ref}$) the cavity. Since the Purcell effect only enhances the transition rate of the resonant, radiative decay channel, the value for the effective Purcell factor can be calculated as \cite[p.~284]{Novotny2012}
    \begin{equation}
		F_\text{P, eff} = \left(\frac{\tau_\text{ref}}{\tau_\text{cav}} - 1\right)\cdot \frac{1}{\eta_\text{QE}} \quad ,
    \end{equation}
    where $\eta_\text{QE}$ denotes the quantum efficiency, which is mainly determined by the blinking of the emitters. Therefore, we use $\eta_\text{QE}\approx\num{0.95}$.   
     
    As mentioned above, due to the deterministic preselection and the flexible cavity design, we were able to directly compare the cavity and free-space emission cases.
    For both setups, we performed time-resolved single-photon counting. Figure~\ref{fig4} displays the results for one measurement of QD A at $L_\text{meas}$ (compare Fig.~\ref{fig3}~(a)). The direct comparison between cavity and
    reference decay is shown. Mono-exponential functions of the form ${f(t) = A\exp{(-t/\tau_\text{dec})} + c_\text{bg}}$ were used to fit the decays. This fit function considers the normalization factor $A$, the decay time constant $\tau_\text{dec}$ and the constant background $c_\text{bg}$. The fits were restricted to the decay region minimizing the fitting error on the decay time constant.
    From the fits, the decay times can be extracted. The values for several QDs are given in Tab.~\ref{tab:decay}.
        \begin{table}
    	\caption{\label{tab:decay}Results of multiple time-resolved single-photon counting measurements. The decay times are extracted by mono-exponential fits. }
    	\begin{ruledtabular}
    		\begin{tabular}{lccccc}
    			QD\# & $\lambda (\si{\nano\meter})$ & $\tau_\text{ref}$ (\si{\nano\second}) & $\tau_\text{cav}$ (\si{\nano\second})& $\tau_\text{ref
    			}/\tau_\text{cav}$ & $F_\text{P, eff}$\\ \hline
    			QD A$^*$ & \num{1306.2} & $\num{1.007(3)}$ & $\num{0.409(3)}$ & $\num{2.46(2)}$ & \num{1.54(2)}\\
    			QD B & \num{1305.0} & $\num{0.632(2)}$ & $\num{0.433(1)}$ & $\num{1.460(6)}$ & \num{0.484(6)}\\
    			QD C & \num{1285.9} &$\num{0.821(2)}$ & $\num{0.521(1)}$ & $\num{1.575(5)}$ & \num{0.605(5)}
    		\end{tabular}
    	\end{ruledtabular}
    	\begin{flushleft}
    		{\footnotesize $^*$ shown in Fig.~\ref{fig4}}
    	\end{flushleft}
    	
    \end{table}
    The comparison between reference and cavity measurement leads to the estimation of the Purcell enhancement. 
    In our
    measurements, we find a maximal reduction of the decay time of up to a factor of $\num{2.46(2)}$, corresponding to $F_\text{P, eff}=\num{1.54(2)}$.
    
    Due to the flexible cavity design, it was possible to investigate emitters distributed over a rather broad spectral region. The lifetime shortening differed for different emitters and also for different measurements at the same emitter. The latter was mainly caused by altering noise conditions. Usually, a more stable cavity resulted in a larger shortening of the decay time. Therefore, we did an analytical description of this situation where the emitter-cavity interaction is described by the overlap of the emitter's and cavity's density of states. These are characterized via their corresponding full widths at half maximum in frequency units $\Delta \nu_\text{cav}$ and $\Delta
    \nu_\text{em}$. The cavity linewidth is determined via the finesse as $\Delta \nu_\text{cav}(\mathcal{F}) = c/(2\,L_\text{eff}\mathcal{F})$.
    In our approach, the emitter properties of the QDs are mainly fixed after the epitaxial growth. On the other hand, the open-cavity design is maximally tunable. Consequently, we particularly considered the influence of the cavity conditions on the Purcell enhancement in a theoretical approach.

    The ideal Purcell factor of a cavity, considering the finite cavity decay rate but excluding vibrational noise, can be calculated as \cite{Auffeves2010, Zifkin2023}
    \begin{equation}\label{eq:FP-ideal}
  		F_\text{P, ideal}=\frac{R_0\Delta\nu_\text{cav}}{\gamma_0(R_0 + \Delta\nu_\text{cav})}\quad ,
    \end{equation}
    with an effective emitter-cavity coupling rate
    \begin{equation}\label{eq:FP-ideal-cases}
	    \begin{split}
    		&R_0 = \frac{3\lambda^3\gamma_0}{2 n_\mathrm{GaAs}^3 \pi^2 w_0^2\,L_\text{eff}}\,\frac{2\,c/\lambda}{\pi(\Delta \nu_\text{cav}(\mathcal{F})+\Delta \nu_\text{em})}\\
    		&=\begin{cases}
    			\frac{6\lambda^2}{n_\mathrm{GaAs}^3 \pi^3 w_0^2}\mathcal{F}\gamma_0,  &\text{ if } \Delta \nu_\text{cav}(\mathcal{F
    			})\gg\Delta \nu_\text{em}\\ \ \\
    			\frac{3\lambda^2c}{n_\mathrm{GaAs}^3 \pi^3 w_0^2 L_\text{eff}\Delta\nu_\text{em}} \,\gamma_0, &\text{ if }\Delta \nu_\text{em}\gg \Delta
    			\nu_\text{cav}(\mathcal{F})
    		\end{cases}
    	\end{split}
    \end{equation}
    where $\lambda$ is the wavelength, $\mathcal{F}$ the cavity finesse, $w_0$ the beam waist of the Gaussian cavity mode, $c$ the speed of light,
    $\gamma_0$ the free space emission rate of the emitter and $n_\mathrm{GaAs}$ the refractive index of the membrane material (GaAs). 
    All included variables beside the emitter linewidth are either design
    parameters or directly determined by them. Additionally, the bad cavity regime (${\Delta \nu_\text{cav}\gg \Delta\nu_\text{em}}$) and the bad emitter regime 
    (${\Delta\nu_\text{em}\gg \Delta\nu_\text{cav}}$) are distinguished. In the bad cavity regime, where ${\Delta \nu_\text{cav}\gg R_0}$, the ideal Purcell factor can be 
    approximated as ${F_\text{P, ideal}\approx R_0/\gamma_0}$ and the typical linear dependence on the finesse follows \cite{Janitz2015}.
    
	Still, this estimation assumes entirely stable cavity conditions. In
    reality, however, fluctuations of the cavity resonance frequency are present (see Fig.~\ref{fig2}~(b) and discussion). These fluctuations of the cavity length lead to a decreased spectral overlap between emitter and cavity mode, which causes a decrease of the ideal Purcell enhancement.
    
    In order to obtain a more general expression, we evaluate the mode overlap in presence of an rms-jitter $\sigma$. 
    Assuming a Gaussian distribution, the derivation of the Purcell factor can be reevaluated (see Appendix \ref{sec:purcell-derivation}) which yields
     \begin{equation}\label{eq:FP-eff}
    		F_\text{P, eff}= \frac{R_\sigma\Delta\nu_\text{cav}}{\gamma_0(R_\sigma + \Delta\nu_\text{cav})}\quad ,
    \end{equation}
    where 
    \begin{equation}
    \begin{split}
    	R_\sigma=&\frac{3\lambda^3  \gamma_0}{2 n_\mathrm{GaAs}^3 \pi^2 w_0^2}\\
    		&\cdot\frac{1}{\sqrt{2\pi}\,\sigma}\cdot\mathrm{exp}\left(\frac{(\Delta \nu_\text{cav}(\mathcal{F})+\Delta
    			\nu_\text{em})^2}{8\left(\sigma c/(L_\text{eff}\lambda)\right)^2}\right)\\
    		&\cdot \mathrm{erfc} \left(\frac{\Delta \nu_\text{cav}(\mathcal{F})+\Delta \nu_\text{em}}{2\sqrt{2}\,\sigma c/(L_\text{eff}\lambda)}\right)\, .
   	\end{split}
    \end{equation}
    For the bad cavity regime ($\Delta \nu_\text{cav}\gg\Delta \nu_\text{em}$) the result given in \cite{Fontana2021,Pallmann2023} is reproduced. The expression here extends the description  of a fluctuating cavity to a full picture in both regimes.\\
    
   Due to the enhanced light-matter interaction and well collectible mode in the cavity, the emission efficiency is typically increased by the cavity which enabled in the past the detection of record values of emission rates \cite{Tomm2021,Nawrath2023,Ding2023}. In order to quantify the brightness, we analyzed the different parts of our setup (for more details on the following see Appendix \ref{sec:brightness}). The total efficiency $\eta_\text{tot}$ is a composite of the excitation efficiency $\eta_\text{exc}$, the quantum efficiency $\eta_\text{QE}$, the influence of the cavity $\eta_\text{cavity}$ as well as both the setup $\eta_\text{setup}$ and the detection efficiency $\eta_\text{det}$:
	\begin{equation}
		\eta_\text{tot} = \eta_\text{exc} \cdot\eta_\text{QE} \cdot \eta_\text{cavity} \cdot \eta_\text{setup} \cdot \eta_\text{det} \quad .
	\end{equation}
	We measured $\eta_\text{setup} = \num{0.23}$ and $\eta_\text{det} = \num{0.77}$. For the cavity, calculations yield $\eta_\text{cavity} = \num{0.11}$ which comprises the probability of an emission into the cavity mode, the transmission through the cavity mirror and the mode matching into the single mode fiber. These depend on the Purcell factor $F_\text{P, eff}$, the transmission ratio through the fiber mirror $T_\text{fiber} / \mathcal{L}_\text{tot}$ and the fiber profile $RC_\text{fiber}$ respectively. Among these experimentally determined factors, the most significant losses occur due to the high losses on the semiconductor side compared to the transmission into the fiber. Additionally, the quantum efficiency determined by the blinking behavior $\eta_\text{QE}\approx\num{0.95}$ has to be considered. Due to the above-band pumping, a clear estimation of $\eta_\text{exc}$ is not possible. Therefore, we find $\eta_\text{tot} = \eta_\text{exc} \cdot\num{0.019}$.
	
	For pulsed excitation (\SI{76}{\mega\hertz}), we found for the studied transition of QD A a maximum countrate at the detectors of $\SI{80}{\kilo\hertz}$. Therefore, we estimate $\eta_\text{tot, meas, max} \approx \num{0.001}$ and consider a small excitation efficiency of $\eta_\text{exc}\leq\num{0.06}$ in order to be consistent in our evaluation. This is reasonable, since the influences of the above band pumping cannot be precisely modeled and are source of a significant uncertainty in this loss estimation. In perspective, resonant excitation schemes could be used to determine the influence of the excitation in more detail.\\
    
	\begin{figure}
		\includegraphics[width=\linewidth]{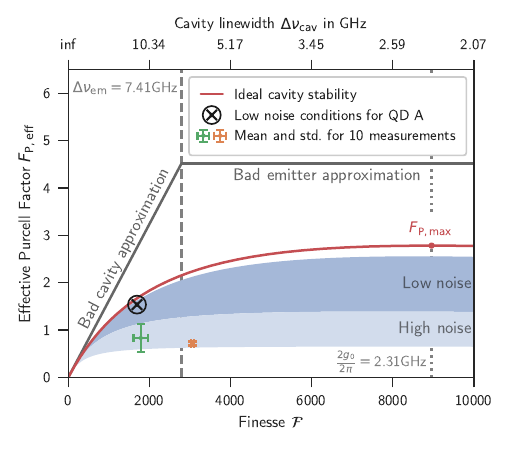}
		\caption{The effective Purcell factor $F_\text{P, eff}$ in dependence of the cavity finesse. Via Eq.~\eqref{eq:FP-ideal-cases}, both bad cavity and bad emitter approximation as $F_\text{P, ideal}\approx R_0/\gamma_0$ are indicated accordingly for QD A (solid grey line). The curve not considering any noise (solid red line) marks the upper limit ($\sigma=0$, full term in Eq.~\eqref{eq:FP-ideal}) whereas the blue areas indicate the experimental conditions corresponding to $\sigma = \SIrange{56}{850}{\pico\meter}$ (via Eq.~\eqref{eq:FP-eff}). Additionally the result from Fig.~\ref{fig4} is displayed (black cross in circle, error bars are too small to be visible) as well as mean results for both fibers (green and orange crosses).}
		\label{fig5}
	\end{figure}
	
    Figure~\ref{fig5} summarizes our obtained results. The effective Purcell enhancement in dependence on the finesse is displayed. Using our experimental parameters, both the bad cavity and the bad emitter approximation ${F_\text{P, ideal}\approx R_0/\gamma_0}$ (solid grey line) are given according to Eq.~\eqref{eq:FP-ideal-cases}. The case of a cavity without any noise ($\sigma=0$) is indicated by the solid red line, whereas the shaded blue areas indicate the experimentally accessible region for low noise ($\sigma_\text{min} = \SI{56}{\pico\meter}$ to $\sigma_\text{typ} = \SI{300}{\pico\meter}$) and high noise ($\sigma_\text{typ}$ to $\sigma_\text{max}=\SI{850}{\pico\meter}$) according to Eq.~\eqref{eq:FP-eff}. The transition from bad cavity to bad emitter regime and the maximum of the effective Purcell enhancement are indicated by vertical lines at $\Delta\nu_\text{em}$ and the emitter-cavity coupling rate $2 g_0$ (calculated from the free-space emission rate, see Appendix \ref{sec:purcell-derivation}).
    
    In addition to the results of the calculations, measured Purcell enhancements and finesses are shown. For fiber~1 with a mean finesse of $\num{1788}\pm\num{179}$, we find a mean Purcell enhancement of $\num{0.83(31)}$ (green cross) and for fiber~2 with finesse $\num{3062}\pm\num{47}$, we find $\num{0.72(4)}$ (orange cross). The mean and standard deviation are derived from \num{10} measurements of different transitions of QDs A, B and C. The measured value for the maximal Purcell enhancement (see result of Fig.~\ref{fig4}) is plotted separately (black cross in circle). 
    Overall, we observe good agreement between the calculated and measured Purcell enhancement, with minor deviations. These can be attributed to three effects not considered in the theoretical description. First, an imperfect dipole overlap can reduce the Purcell enhancement (included by an overlap factor $\xi$ in Appendix \ref{sec:purcell-derivation}). This effect becomes increasingly important as the cavity linewidth becomes more narrow and a significant polarization splitting of the cavity mode is observed. Second, the QD's position inside the membrane may deviate from the maximum field intensity. Third, the reference decay time is already shortened by the QD's placement in front of the bottom DBR. Depending on the distance between emitter and DBR, the expected decay time is already reduced by a factor of up to \num{1.07} compared to emission in GaAs bulk material (see Appendix \ref{sec:DBR}). All three effects are not included in the simulation shown, but could be easily incorporated if the respective parameters were determined.\\

    In summary, we demonstrated an open fiber-cavity working with QDs emitting in the telecom O-band. Due to the flexible design and the deterministic preselection we were able to compare the emission of several emitters inside and outside the cavity. We measured a reduction of the decay time inside the cavity of up to $\num{2.46(2)}$ due to Purcell enhancement. 
    
    Current limitations of our system are the emitter linewidth, the maximal achievable finesse, the minimal achievable cavity length and the cavity stability. Firstly, a reduction of the emitter linewidth would lead to a linear increase of the achievable effective Purcell factor, provided that the cavity linewidth can stay equally small. A narrower emitter linewidth could potentially be achieved by using a gated QD structure. Future implementations may also include the realization of resonant or quasi-resonant excitation schemes. Secondly, the maximal finesse is currently limited by the surface roughness and surface defects due to the advanced MOVPE growth of the QDs. A possible improvement could be realized by surface passivation \cite{Guha2017, Najer2019}. Thirdly, the minimal achievable cavity length is limited by the profile depth, the fiber tilt and the penetration depth into the mirrors. A reduction of any of these factors could enable shorter cavity lengths, which would increase the Purcell enhancement in the bad emitter regime according to Eq.~\eqref{eq:FP-ideal-cases}. Likewise, the cavity linewidth (top x-axis in Fig.~\ref{fig5}) is larger for shorter cavity lengths and therefore the transition into the bad emitter regime occurs at higher finesse. Lastly, the fluctuations of the cavity due to mechanical vibrations pose a challenge. Our study contains a consistent analysis which shows that for an optimized system the noise is currently not a major limiting factor. It is conceivable, that our in-situ noise analysis enables further improvements in handling the fluctuations i.e. by post-selection or active locking. In total, the presented system can be a promising building block for quantum technological applications. 
    
    \section*{Acknowledgements}
    The authors acknowledge the support provided by T.~ Herzog and J.~Wecker in early stages of the experiment and their previous works. We also thank R.~Sittig for the sample preparation. This work directly benefited from the knowledge base generated in the Karlsruhe School of Optics and Photonics (KSOP) and the QR.X project. The work is financially supported by the Baden-Württemberg Foundation and the Ministry of Science, Research and Arts Baden-Württemberg (MWK) via the project `Telecom SPS'.

    \appendix
     \section{Estimation of the Brightness}\label{sec:brightness}
    As enumerated above, the total collection efficiency consists of five factors
    \begin{equation}
    	\eta_\text{tot} = \eta_\text{exc} \cdot\eta_\text{QE} \cdot \eta_\text{cavity} \cdot \eta_\text{setup} \cdot \eta_\text{det}
    \end{equation}
    the excitation efficiency $\eta_\text{exc}$, the quantum efficiency $\eta_\text{QE}$, the influence of the cavity $\eta_\text{cavity}$, the setup $\eta_\text{setup}$ and the detection efficiency $\eta_\text{det}$. Among these, the influences of the cavity can be again split into another three factors
    \begin{equation}
    	\eta_\text{cavity} = \eta_\text{mode}\cdot\eta_\text{trans}\cdot \eta_\text{fib}\, ,
    \end{equation}
     which describe the percentage of emission into the cavity mode, the transmission rate through the fiber mirror and the coupling efficiency into the single mode fiber respectively.\\
    Therefore, the total collection efficiency is 
    \begin{equation}
    	\eta_\text{tot} = \eta_\text{exc} \cdot\eta_\text{QE} \cdot \eta_\text{mode}\cdot\eta_\text{trans}\cdot \eta_\text{fib} \cdot \eta_\text{setup} \cdot \eta_\text{det}
    \end{equation}
    and the determined values of the individual components are listed in table \ref{tab:losses}.
    
    First, the detection efficiency is specified by the supplier of the detector as 
    \begin{equation}
    	\eta_\text{det} = \num{0.77}\,.
    \end{equation}
    Second, the setup efficiency was measured from the inner side of the cryo feedthrough to the detector input, comprising a fiber connection over multiple levels of our building, to be 
    \begin{equation}
    	\eta_\text{setup} = \num{0.23}\,.
    \end{equation}
    Third, the emission into the cavity mode (also called $\beta$-factor) is given by \cite[p.~204]{Fox2006}
    \begin{equation}
    	\eta_\text{mode} =\beta =\frac{\gamma_\text{cav}}{\gamma_0 + \gamma_\text{cav}}= \frac{F_\text{P, eff}}{1 + F_\text{P, eff}}\, ,
    \end{equation}
    where $\gamma_\text{cav}$ and $\gamma_0$ are the emission rate into the cavity mode and the free space emission rate. $F_\text{P, eff}$ is the effective Purcell factor as it is defined in Appendix \ref{sec:purcell-derivation}.
    With $F_\text{P, meas}=\num{1.54}$ we obtain 
    \begin{equation}
    	\eta_\text{mode} =\num{0.61}\,.
    \end{equation}
    Fourth, the transmission rate through the fiber mirror is given by the ratio of the losses transmitted through the fiber mirror over the total losses of the cavity
    \begin{equation}
    	\eta_\text{trans} = \frac{T_\text{fib}}{\mathcal{L}_\text{tot}}\,.
    \end{equation}
    With our experiment results $T_\text{fib} = \SI{1000}{ppm}$ and $\mathcal{L}_\text{tot} = \mathcal{L}_\text{tot}^\text{fib}+ \mathcal{L}_\text{tot}^\text{sc} =  \SI{1010}{ppm}+ \SI{1864}{ppm} = \SI{2874}{ppm}$ the transmission rate is
    \begin{equation}
    	\eta_\text{trans} = \num{0.35}\,.
    \end{equation}
    This ratio can be significantly improved by a lower transmission mirror on the semiconductor side and the use of the higher transmissive mirror on the detection side, i.e. the fiber side.\\
    Fifth, the last contributing factor of the cavity is the mode matching efficiency into the fiber mode, which is given by \cite[Eq.~10]{Hunger2010}
    \begin{equation}
    	\eta_\text{fib} = \frac{4}{\left(\frac{w_\text{f}}{w_\text{m}} + \frac{w_\text{m}}{w_\text{f}}\right)^2 +
    		\left(\frac{\pi n_\text{f}w_\text{f}w_\text{m}}{\lambda\, RC}\right)^2}\, ,
    \end{equation}
    where $w_m = \SI{2.50}{\micro\meter}$ is the mode waist at the fiber mirror, $w_f = \SI{4.8}{\micro\meter}$ is the radius of the mode field in the single mode fiber specified by the manufacturer, $n_f=\num{1.45}$ is the refractive index of the fiber material specified by the manufacturer and $RC=\SI{34.3}{\micro\meter}$ is the measured radius of curvature of fiber~1. With these parameters
    \begin{equation}
    	\eta_\text{fib} = \num{0.54}\, ,
    \end{equation}
    leading to a total cavity extraction efficiency of 
    \begin{equation}
    	\eta_\text{cav} = \num{0.11}\, .
    \end{equation}
    Sixth, the quantum efficiency $\eta_\text{QE}$ equals the probability of the excited state decaying into the observed radiative decay channel.
    From the blinking we observed in the $g^{(2)}$-measurement, we expect the quantum dot to recombine radiatively in $\SI{95}{\percent}$ of all cases, such that
    \begin{equation}
    	\eta_\text{QE} = \num{0.95}\,.
    \end{equation}
    Due to the above band gap excitation, a reasonable estimation of the excitation efficiency of our excited state is not possible. Therefore, we 
    compare our deduced efficiencies with our measured photon rate. We measured a maximum photon count rate of $\SI{80}{\kilo\hertz}$ in our experiment, which
    equals a total extraction efficiency of 
    \begin{equation}
    	\eta_\text{tot,meas,max} = \frac{\SI{80}{\kilo\hertz}}{\SI{76}{\mega\hertz}} = \num{0.0011}\, .
    \end{equation}
    From that we can calculate the remaining seventh factor, the excitation efficiency, to be
    \begin{equation}
    	\eta_\text{exc}=\num{0.06}\,.
    \end{equation}
    \begin{table}
    	\caption{\label{tab:losses}Extraction efficiency broke down by individual loss sources.}
    	\begin{ruledtabular}
    		\begin{tabular}{ccc}
    			& Eff.        & Source       \\ \hline
    			$\eta_\text{QE}$    & \num{0.95}  & meas. \\
    			$\eta_\text{mode}$  & \num{0.61}  & calc. from meas. \\
    			$\eta_\text{trans}$ & \num{0.35}  & manufacturer and meas. \\
    			$\eta_\text{fib}$   & \num{0.54}  & calc. from meas.\\
    			$\eta_\text{setup}$ & \num{0.23}  & meas. \\
    			$\eta_\text{det.}$  & \num{0.77}  & manufacturer \\ \hline
    			$\eta_\text{tot}$   & \num{0.019} &
    		\end{tabular}
    	\end{ruledtabular}
    \end{table}
    \section{Impedance Matching}\label{sec:impedance-matching}
    In this section we like to compare the depth of our resonance dip with calculations for the impedance matching of our cavity.
    The contrast of the resonance dip is given by the observed loss channel $\mathcal{L}^\text{fib}_\text{trans}$ in comparison to all other loss channels $\mathcal{L}^\text{fib}_\text{tot}$ and $\mathcal{L}^\text{sc}_\text{tot}$. It can be calculated as \cite[Eq.~19]{Gallego2016}
    \begin{equation}
    	C_\text{imp} = 
    	1 - \frac{\left(2 \mathcal{L}^\text{fib}_\text{trans} - \mathcal{L}^\text{fib}_\text{tot} -
    		\mathcal{L}^\text{sc}_\text{tot}\right)^2}{\left(\mathcal{L}^\text{fib}_\text{tot}  + \mathcal{L}^\text{sc}_
    		\text{tot}\right)^2}
    	=\num{0.908}\, ,
    \end{equation}
    where $\mathcal{L}^\text{fib}_\text{trans}=\SI{1000}{ppm}$ is the transmission through the fiber mirror, $\mathcal{L}^\text{fib}_\text{tot}=\SI{1010}{ppm}$ the total losses on the fiber side and $\mathcal{L}^\text{sc}_\text{tot}=\SI{1863}{ppm}$ the total losses on the semiconductor side. \\
    This contrast is in excellent accordance to our measured contrast of
    \begin{equation}
    	C_\text{imp, meas} = 0.912\,.
    \end{equation}
    In perspective this measurement can be an easy way to determine the loss ratios of the cavity and we showed in addition to the other measurements that our estimations of the losses are reasonable and congruent.
    
    \section{Derivation of the effective Purcell factor}\label{sec:purcell-derivation}
    In this section we like to present how we derived the equations for the effective Purcell factor (Eq. \eqref{eq:FP-ideal} and
    \eqref{eq:FP-eff} in the main text).
    The effective Purcell factor is defined as the fraction of the rate of emission from the cavity mode $\gamma_\text{cav}$
    over the free space emission rate $\gamma_0$.
    The effective cavity loss rate can be calculated as \citep{Auffeves2010, Zifkin2023}
    \begin{equation}
    	\gamma_\text{cav} = \frac{R \Delta\omega_\text{cav}}{R + \Delta\omega_\text{cav}} ,
    \end{equation}
    where $\Delta\omega_\text{cav}$ equals the cavity linewidth and $R$ is the effective coupling rate between the emitter and the cavity. The latter depend on the density of states of the emitter and the cavity and can be calculated using Fermi's
    golden rule
    \begin{equation}
       	\label{eq:cavity-rate}
		R = \frac{2\pi}{\hbar^2}\xi^2\int_0^{\infty}\left|M_{12}\right|^2 g(\omega) \Lambda(\omega) \,\text{d}\omega
    \end{equation}
    and, averaged over all possible dipole orientations, the free space emission rate is
    \begin{equation}
    	\label{eq:emitter-rate}
    	\gamma_0 = \frac{2\pi}{\hbar^2}\frac{1}{3}\left|M_{12}\right|^2 g(\omega)
    	=\frac{1}{3}\frac{\mu_{12}^2 \omega_\text{em}^{3}}{\hbar\epsilon_0 \pi c^3}\, ,
    \end{equation}
    where $\xi^2$ is the dipole overlap in the cavity, $\left|M_{12}\right|^2=\mu_{12}^2\hbar\omega / (2\epsilon_0
    V_0)$ is the transition matrix element, $g(\omega)$ the desity of states of the
    emitter, equal to $g(\omega)=\omega_\text{em}^{2}V_0 / (\pi^2 c^3)$ in free space, and $\Lambda(\omega)$ the
    density of states of the cavity \cite[pp.~201-203]{Fox2006}.
    If the cavity and emitter resonance are displaced by $\delta$, with a propability density function of this
    displacement $PDF(\delta)$, the cavity emission rate becomes
    \begin{align}
    	\label{eq:cavity-rate-displaced}
    	R =& \frac{2\pi}{\hbar^2}\xi^2\\
    	&\cdot\int_0^{\infty} \left(\int_{-\infty}^{\infty}\left|M_{12}\right|^2
    	\, g(\omega)\, \Lambda(\omega+\delta)\, PDF(\delta)\,\text{d}\delta\right) \,\text{d}\omega\,.\notag
   	\end{align}
    For a perfectly stable cavity on the emitter resonance, the $PDF(\delta)=\delta_\text{Dirac}(\delta)$
    (where $\delta_\text{Dirac}$ denotes the Dirac delta function)
    and Eq. \eqref{eq:cavity-rate-displaced} equals Eq. \eqref{eq:cavity-rate}.
    
    However, in our calculations we model the cavity instability by a Gaussian distributed jitter
    \begin{equation}
    	\label{eq:cavity-displacement-PDF}
    	PDF(\delta) =\frac{1}{\sqrt {2\pi}\sigma_{\omega}} \exp\left(-\frac{\delta^2}{2\sigma_{\omega}^2}\right),
    \end{equation}
    where $\sigma_{\omega}$ denotes the rms-deviation of the line jitter $\delta$ in angular frequency units (or in
    general in the same units as $\delta$ and $\omega$).
    
    Inserting these definitions into the fraction to calculate the effective Purcell factor leads to
    	\begin{align}
    			\label{eq:effective-purcell}
    			F_{\text{P, eff}}&=\frac{\gamma_\text{cav}}{\gamma_0}\\ \notag
    			&=\frac{2\pi}{\hbar^2} \xi^2 \frac{3\hbar\epsilon_0 \pi
    				c^3}{\mu_{12}^2\omega_\text{em}^{3}} \frac{\Delta\omega_\text{cav}}{R + \Delta\omega_\text{cav}}\\ \notag
    			&\cdot \int_0^{\infty} \left(\int_{-\infty}^{\infty} \left|M_{12}\right|^2\,
    			g(\omega)\, \Lambda(\omega+\delta)\, PDF(\delta)\,\text{d}\delta\right) \,\text{d}\omega\\\notag
    			&=\frac{2\pi}{\hbar^2} \xi^2 \frac{3\hbar\epsilon_0 \pi c^3}{\mu_{12}^2 \omega_\text{em}^{3}} 
    			\frac{\mu_{12}^2\hbar}{2\epsilon_0} \frac{\Delta\omega_\text{cav}}{R + \Delta\omega_\text{cav}}\\\notag
    			&\cdot\int_0^{\infty}\left(
    			\int_{-\infty}^{\infty}\frac{\omega}{V(\omega)}\, g(\omega)\, \Lambda(\omega+\delta)\, PDF(\delta)\,\text{d}\delta\right) \,\text{d}\omega\,\\ \notag
    			&=F_\text{prefactor} \frac{\Delta\omega_\text{cav}}{R + \Delta\omega_\text{cav}} I.
    	\end{align}
    It is noteworthy, that the coupling strength, or rather the mode volume $V(\omega)$ in general depends on the
    frequency $\omega$.
    However, if we shift the integral to the resonance frequency $\omega = \omega_0 + \widetilde{\omega}$
    and replace the integration variable by $\widetilde{\omega}=\omega-\omega_0$, the integral is
    \begin{equation}
    	\int_{-\omega_0}^{\infty}\left(\int_{-\infty}^{\infty}\frac{\omega_0+\widetilde{\omega}}{V(\omega_0
    		+\widetilde{\omega})}\, g(\widetilde{\omega})\, \Lambda(\widetilde{\omega}+\delta)\, PDF(\delta)\,\text{d}\delta\right) \,\text{d}\widetilde{\omega}\,.
    \end{equation}
    The cavity's mode volume $V(\omega_0+\widetilde{\omega})$ is almost constant if the resonance frequency $\omega_0$
    is much larger than the integration over $\widetilde{\omega}$, i.e. $\omega_0\gg\widetilde{\omega}$.
    On the one hand, using this approximation we can extract the constant $V(\omega_0+\widetilde{\omega})\approx V(\omega_0)= V_\text{cav}$
    from the integral and, on the other hand, we can extend the lower bound of the integral over $\widetilde{\omega}$ to $-\omega_0 \rightarrow-\infty$.
    The integral left to be solved is
    \begin{equation}
    	I= \int_{-\infty}^{\infty}\left(\int_{-\infty}^{\infty}\left(\omega_0+\widetilde{\omega}\right)\, g(\widetilde{\omega}) \Lambda(\widetilde{\omega}+\delta)PDF(\delta)\,\text{d}\delta\right) \,\text{d}\widetilde{\omega}
    \end{equation}
    and the prefactor of the integral (see Eq. \eqref{eq:effective-purcell}) is
    \begin{equation}
    \begin{split}
    	\label{eq:purcell-prefactor}
    	F_{\text{prefactor}}&=\xi^2\frac{3 \pi^2 c^3}{\omega_\text{em}^{3}V_\text{cav}}\\
    	&=\xi^2\frac{3 \pi^2 \lambda^3}{(2\pi)^{3} n^3 V_\text{cav}}\\
    	&=\xi^2\frac{3\lambda^3}{2 n^3 \pi^2 w_0^2 L_\text{eff}}\,.
   	\end{split}
    \end{equation}
    Here the cavity's mode volume is given by the Gaussian mode waist $w_0$ in the semiconductor membrane and the effective cavity length $L_\text{eff}$ as $V_\text{cav}=1/4\, \pi w_0^2 L_\text{eff}$. The mode waist in the membrane can be calculated as \cite[Eq.~20]{Dam2018}
    \begin{equation}
    \begin{split}
    	\label{eq:mode-waist}
    	w_0 = &\Bigg( \frac{\lambda}{\pi}\bigg[ \left(L_\text{air}+L_\text{mem}/n_\text{GaAs}\right)\cdot RC\bigg.\Bigg. \\
    	&\Bigg.\bigg. -\left(L_\text{air}+L_\text{mem}/n_\text{GaAs}\right)^2\bigg]^\frac{1}{2} \Bigg)^\frac{1}{2}\, ,
    \end{split}
    \end{equation}
    where $RC$, $L_\text{eff}$, $L_\text{air}$ and $L_\text{mem}$ equal the radius of curvature of the fiber mirror, the effective energy distribution length, the length of the air gap and the membrane thicknesses as defined in the main text. Also $R$ is proportional to the integral $I$, with the prefactor $F_\text{prefactor}$
    \begin{equation}
    \label{eq:em_cav_coupling}
    R= \frac{2\pi}{\hbar^2}\xi^2\frac{\mu_{12}^2\hbar}{2\epsilon_0 V_\text{cav}} I = \xi^2\frac{3 \pi^2 c^3}{\omega_\text{em}^3 V_\text{cav}} \gamma_0 I = F_\text{prefactor} \gamma_0 I
    \end{equation}
    The density of states are represented by normalized Loretzian functions \cite[p.~202]{Fox2006}
    \begin{equation}
    	g(\widetilde{\omega})=\frac{1}{\pi \Delta\omega_\text{em}/2}\,\frac{(\Delta\omega_\text{em}/2)^2}{\widetilde{\omega}^2+(\Delta\omega_
    		\text{em}/2)^2}
    \end{equation}
    and
    \begin{equation}
    	\Lambda(\widetilde{\omega})=\frac{1}{\pi \Delta\omega_\text{cav}/2}\,\frac{(\Delta\omega_\text{cav}/2)^2}{\widetilde{\omega}^2+(
    		\Delta\omega_\text{cav}/2)^2}\, ,
    \end{equation}
    where $\Delta\omega_\text{cav}$ and $\Delta\omega_\text{em}$ equal the full width at half maximum of the cavity and emitter line respectively.
    
    With this the solution of the integral
    \begin{equation}
    \label{eq:imperfect-integral}
    \begin{split}
    	I = &\frac{\omega_0}{\sqrt{2\pi}\sigma_{\omega}}\\
    	&\cdot\exp\left( \frac{(\Delta\omega_\text{cav}+\Delta\omega_\text{em})^2}{4\cdot 2\sigma_{\omega}^2}\right)\\
    	&\cdot\text{erfc} \left(\frac{\Delta\omega_\text{cav}+\Delta\omega_\text{em}}{2\sqrt{2}\sigma_{\omega}}\right)
    \end{split}
    \end{equation}
    and for a perfectly stable cavity
    \begin{equation}
    	\label{eq:perfect-integral}
    	I_{\delta=0} = \frac{2\omega_0}{\pi (\Delta\omega_\text{cav}+\Delta\omega_\text{em})}\,.
    \end{equation}
    It can become handy to rewrite these solutions either in ordinary frequency units with the resonance frequency $\nu_0$, 
    the resonance linewidths $\Delta\nu_\text{cav}$ and $\Delta\nu_\text{em}$ and the line jitter $\sigma_\nu$, or
    in spatial units by expressing the Lorentzian resonance function in terms of a cavity length changes $\Delta L_\text{cav}$,
    $\Delta L_\text{em}$ and the spatial line jitter $\sigma$. For the former the result is
    \begin{equation}
        \begin{split}
    	I_\nu =& \frac{\nu_0}{\sqrt{2\pi}\sigma_{\nu}}\\
    	&\cdot\exp\left( \frac{(\Delta\nu_\text{cav}+\Delta\nu_\text{em})^2}{4\cdot 2\sigma_{\nu}^2}\right)\\
    	&\cdot\text{erfc} \left(\frac{\Delta\nu_\text{cav}+\Delta\nu_\text{em}}{2\sqrt{2}\sigma_{\nu}}\right)
    	\end{split}
    \end{equation}
    and
    \begin{equation}
    	I_{\nu,\, \delta=0} = \frac{2\nu_0}{\pi (\Delta\nu_\text{cav}+\Delta\nu_\text{em})}\, ,
    \end{equation}
    whereas for the latter    
    \begin{equation}
    \begin{split}
    	I_L = &\frac{L_\text{eff}}{\sqrt{2\pi}\sigma}\\
    	&\cdot\exp\left( \frac{(\Delta L_\text{cav}+\Delta L_\text{em})^2}{4\cdot 2\sigma^2}\right)\\
    	&\cdot\text{erfc} \left(\frac{\Delta L_\text{cav}+\Delta L_\text{em}}{2\sqrt{2}\sigma}\right)
    \end{split}
    \end{equation}
    and
    \begin{equation}
    	I_{L,\, \delta=0} = \frac{2 L_\text{eff}}{\pi (\Delta L_\text{cav}+\Delta L_\text{em})}\,.
    \end{equation}
    To convert the angular frequencies to ordinary frequencies, one has to divide $\omega_0, \sigma_\omega, \Delta\omega_\text{cav}$ 
    and $\Delta\omega_\text{em}$ by $2\pi$ 
    and to convert the ordinary frequency units to spatial units, one divides $\sigma_\nu, \Delta\nu_\text{cav}$ and $\Delta\nu_\text{em}$ by $c/(L_\text{eff} \lambda)$. 
    Furthermore, $\nu_0=c/\lambda$ and the cavity linewidth can be expressed through the finesse $\mathcal{F}$ 
    as $\Delta\nu_\text{cav} = c/(2L_\text{eff}\mathcal{F})$ or $\Delta L_\text{cav}=\lambda/(2\mathcal{F})$.
    
    \subsection*{Expression in cavity QED parameters}
    Conventionally these solutions are expressed using the angular frequency parameters: $\kappa=\Delta\omega_\text{cav}$ the cavity linewidth (FWHM), 
    $\gamma+\gamma^*=\Delta\omega_\text{em}$ the total emitter linewidth (FWHM) comprising the lifetime limited linewidth $\gamma$ and the pure dephasing rate $\gamma^*$,
    and $g$ the emitter-cavity coupling rate.
    The latter is defined as 
    \begin{equation}
    	g(\rho, z) = g_0 u(\rho,z)
    \end{equation}
    where $g_0 = M_{12}/\hbar$ and $u(\rho,z)$ is the spatial distribution of the mode in cylindrical coordinates (the Gaussian mode distribution).
    One can calculate \cite[pp.~199,~201]{Fox2006}
    \begin{equation}
    	\label{eq:emitter-cavity-coupling-rate}
    	g_0=\sqrt{\frac{\mu_{12}^2\omega_0}{2\epsilon_0\hbar V_\text{cav}}}=\sqrt{\frac{\gamma}{2}\, \frac{3 \lambda^2 c}{n^3\pi^2 w_0^2 L_\text{eff}}}
    \end{equation}
    and identify our $F_\text{prefactor}$ (Eq. \eqref{eq:purcell-prefactor}), $I_{\delta=0}$ (Eq. \eqref{eq:perfect-integral}) and $R_{\delta=0}$ (Eq. \eqref{eq:em_cav_coupling}) as
    \begin{equation}
    \begin{split}
    	F_{\text{prefactor}}&=\xi^2 \frac{g_0^2\lambda}{c\gamma}\\
    	&=\xi^2 \frac{\gamma}{2}\, \frac{3 \lambda^2 c}{n^3 \pi^2 w_0^2 L_\text{eff}}\frac{\lambda}{c\gamma}\\
    	&=\xi^2\frac{3\lambda^3}{2 n^3 \pi^2 w_0^2 L_\text{eff}}\, ,
   	\end{split}
    \end{equation}
    \begin{equation}
    	I_{\delta=0} = \frac{4\pi c/\lambda}{\pi (\kappa+\gamma+\gamma^*)}=\frac{2\omega_0}{\pi(\Delta\omega_\text{cav}+\Delta\omega_\text{em})}\, ,
    \end{equation}
    and
    \begin{equation}
    \begin{split}
    	R_{\delta=0} &= \xi^2\frac{4g_0^2}{\kappa+\gamma+\gamma^*}\\
    	&=\xi^2 \frac{g_0^2\lambda}{c\gamma}\gamma\frac{4\pi c/\lambda}{\pi (\kappa+\gamma+\gamma^*)}\\
    	&=F_{\text{prefactor}}\gamma I_{\delta=0}\, .
   	\end{split}
    \end{equation}
    In total this leads to the well known formula for the effective Purcell factor of a perfectly stable cavity \cite{Auffeves2010, Zifkin2023}
    \begin{equation}
    \begin{split}
    	F_\text{P, ideal} &=\left(\frac{\gamma_\text{cav}}{\gamma_0}\right)_{\delta=0} = \frac{R_{\delta=0}\, \kappa}{\gamma(R_{\delta=0} + \kappa)}\\
    	&=F_{\text{prefactor, } \delta=0}\cdot\frac{\kappa}{R_{\delta=0} + \kappa}\cdot I_{\delta=0}\\
    	&= \xi^2 \frac{4g_0^2}{\gamma (\kappa+\gamma+\gamma^*)}\frac{\kappa}{\xi^2\frac{4g_0^2}{\kappa+\gamma+\gamma^*} + \kappa}
   	\end{split}
    \end{equation}
    and for a Gaussian jitter it leads to
    \begin{equation}
    \begin{split}
    	F_\text{P, eff} &=\frac{R\kappa}{\gamma(R+ \kappa)}\\
    	&= \xi^2 \frac{g_0^2\lambda}{c\gamma} I \frac{\kappa}{\xi^2 \frac{g_0^2\lambda}{c} I+ \kappa}\\
    	&= \xi^2\frac{g_0^2}{\gamma} \frac{2\pi}{\omega_0} I \frac{\kappa}{\xi^2 g_0^2 \frac{2\pi}{\omega_0} I+ \kappa} \, ,
    \end{split}
    \end{equation}
    where,  using Eq. \ref{eq:imperfect-integral},
	\begin{equation}
	\begin{split}
	I=&\frac{\omega_0}{\sqrt{2\pi} \sigma_\omega}\\
		&\cdot\exp\left( \frac{(\kappa+\gamma+\gamma^*)^2}{4\cdot 2\sigma_{\omega}^2}\right)\\
    	&\cdot\text{erfc} \left(\frac{\kappa+\gamma+\gamma^*}{2\sqrt{2}\sigma_{\omega}}\right)\, .
	\end{split}
	\end{equation}	        
    
    Using our experiment parameters (emitter lifetime $\gamma/(2\pi)=\SI{0.16}{\giga\hertz}$, wavelength $\lambda=\SI{1.31}{\micro\meter}$,
    refractive index $n=n_\text{GaAs}=\num{3.41}$, mode waist $w_0=\SI{2.28}{\micro\meter}$ (c.f. Eq. \eqref{eq:mode-waist}) and cavity length $L_\text{eff}=\SI{7.25}{\micro\meter}$)
    in Eq.~\eqref{eq:emitter-cavity-coupling-rate}, we can calculate a theoretical value for the emitter-cavity coupling rate of
    \begin{equation}
    	\frac{g_0}{2\pi} = \SI{1.15}{\giga\hertz}\, ,
    \end{equation}
    indicated in Fig.~\ref{fig5}. 
    \section{Resonance shift for fiber in contact}\label{sec:strainshift}
\begin{figure}[H]
 		 	\includegraphics[width=\columnwidth]{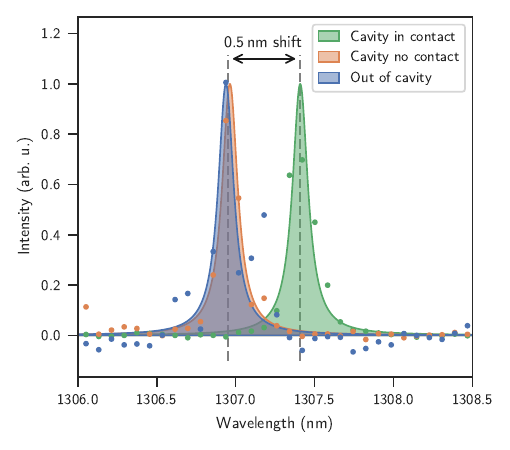}
 		 	\caption{Shift of the emitter resonance, when the fiber tip gets in contact with the quantum dot sample. Three resonance spectra are displayed: When the emitter is out of the cavity (blue) or the fiber is not in contact with the sample (orange) the emission is at \SI{1306.9}{\nano\meter}. When the fiber is in contact with the sample (green), the emission is shifted by \SI{0.5}{\nano\meter} to \SI{1307.4}{\nano\meter}. This shift is most likely due to the induced mechanical strain.}
 		 	\label{fig:strainshift}
	 \end{figure}

    In the main text, we discuss the shift of the resonance frequency due to the contact of the fiber with the quantum dot sample. Fig. \ref{fig:strainshift} provides a detailed view of this effect.
    \section{Lifetime shortening in front of the DBR}\label{sec:DBR}
\begin{figure}[H]
 		 	\includegraphics[width=\columnwidth]{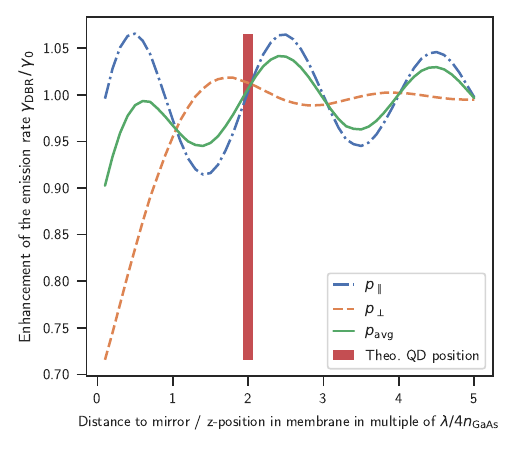}
 		 	\caption{Results of the numeric simulation of the emission rate in front of a 10 layer GaAs/AlAs DBR. The enhancement is dependent on the dipole orientation. For a dipole oriented parallel to the mirror surface the blue dash-dotted line is depicted, for a perpendicular oriented dipole the orange dashed line. For a random orientation the statistical average is plotted as solid green line. The influence at the aimed QD position (red area) is negligible.}
 		 	\label{figDBR}
	 \end{figure}

\begin{figure}[H]
 		 	\includegraphics[width=\columnwidth]{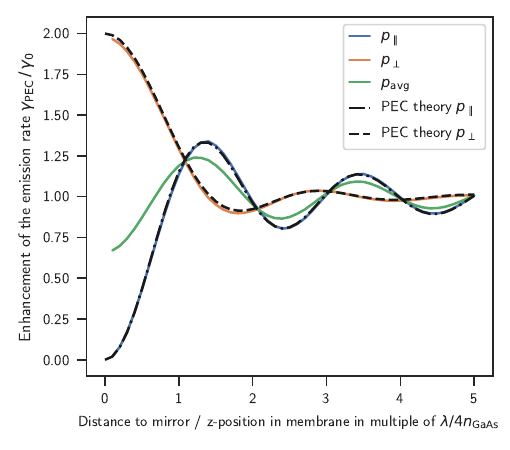}
 		 	\caption{Results of the numeric simulation of the emission rate in front of a perfect electric conductor (PEC). The parallel dipole orientation is indicated by a solid blue line, the perpendicular orientation by a solid orange line and the statistical average by a solid green line. The analytic solutions for a dipole in front of a PEC \cite{FurtakWells2018} are indicated by the dash-dotted and dashed lines and fit our numeric results perfectly.}
 		 	\label{figPEC}
	 \end{figure}

    We estimated the influence of the DBR fabricated below the QDs from a numeric, finite-element simulation using COMSOL~Multiphysics\textsuperscript{\textregistered} . The result of this simulation is depicted in Fig.~\ref{figDBR}. To show that our simulation results are reliable, we performed a benchmark using a dipole in front of a perfect electric conductor (PEC), shown in Fig.~\ref{figPEC}. The results are perfectly matching the analytic predictions for this problem \cite{FurtakWells2018}.\\ \ \\ \ \\
    

    \bibliographystyle{apsrev4-2} 
    \renewcommand{\discretionary}{}

\begin{thebibliography}{48}%
\makeatletter
\providecommand \@ifxundefined [1]{%
 \@ifx{#1\undefined}
}%
\providecommand \@ifnum [1]{%
 \ifnum #1\expandafter \@firstoftwo
 \else \expandafter \@secondoftwo
 \fi
}%
\providecommand \@ifx [1]{%
 \ifx #1\expandafter \@firstoftwo
 \else \expandafter \@secondoftwo
 \fi
}%
\providecommand \natexlab [1]{#1}%
\providecommand \enquote  [1]{``#1''}%
\providecommand \bibnamefont  [1]{#1}%
\providecommand \bibfnamefont [1]{#1}%
\providecommand \citenamefont [1]{#1}%
\providecommand \href@noop [0]{\@secondoftwo}%
\providecommand \href [0]{\begingroup \@sanitize@url \@href}%
\providecommand \@href[1]{\@@startlink{#1}\@@href}%
\providecommand \@@href[1]{\endgroup#1\@@endlink}%
\providecommand \@sanitize@url [0]{\catcode `\\12\catcode `\$12\catcode
  `\&12\catcode `\#12\catcode `\^12\catcode `\_12\catcode `\%12\relax}%
\providecommand \@@startlink[1]{}%
\providecommand \@@endlink[0]{}%
\providecommand \url  [0]{\begingroup\@sanitize@url \@url }%
\providecommand \@url [1]{\endgroup\@href {#1}{\urlprefix }}%
\providecommand \urlprefix  [0]{URL }%
\providecommand \Eprint [0]{\href }%
\providecommand \doibase [0]{http://dx.doi.org/}%
\providecommand \selectlanguage [0]{\@gobble}%
\providecommand \bibinfo  [0]{\@secondoftwo}%
\providecommand \bibfield  [0]{\@secondoftwo}%
\providecommand \translation [1]{[#1]}%
\providecommand \BibitemOpen [0]{}%
\providecommand \bibitemStop [0]{}%
\providecommand \bibitemNoStop [0]{.\EOS\space}%
\providecommand \EOS [0]{\spacefactor3000\relax}%
\providecommand \BibitemShut  [1]{\csname bibitem#1\endcsname}%
\let\auto@bib@innerbib\@empty
\bibitem [{\citenamefont {Gisin}\ and\ \citenamefont {Thew}(2007)}]{Gisin2007}%
  \BibitemOpen
  \bibfield  {author} {\bibinfo {author} {\bibfnamefont {N.}~\bibnamefont
  {Gisin}}\ and\ \bibinfo {author} {\bibfnamefont {R.}~\bibnamefont {Thew}},\
  }\href {\doibase 10.1038/nphoton.2007.22} {\bibfield  {journal} {\bibinfo
  {journal} {Nature Photonics}\ }\textbf {\bibinfo {volume} {1}},\ \bibinfo
  {pages} {165} (\bibinfo {year} {2007})}\BibitemShut {NoStop}%
\bibitem [{\citenamefont {Vajner}\ \emph {et~al.}(2022)\citenamefont {Vajner},
  \citenamefont {Rickert}, \citenamefont {Gao}, \citenamefont {Kaymazlar},\
  and\ \citenamefont {Heindel}}]{Vajner2022}%
  \BibitemOpen
  \bibfield  {author} {\bibinfo {author} {\bibfnamefont {D.~A.}\ \bibnamefont
  {Vajner}}, \bibinfo {author} {\bibfnamefont {L.}~\bibnamefont {Rickert}},
  \bibinfo {author} {\bibfnamefont {T.}~\bibnamefont {Gao}}, \bibinfo {author}
  {\bibfnamefont {K.}~\bibnamefont {Kaymazlar}}, \ and\ \bibinfo {author}
  {\bibfnamefont {T.}~\bibnamefont {Heindel}},\ }\href {\doibase  10.1002/qute.202100116} {\bibfield  {journal} {\bibinfo  {journal} {Advanced
  Quantum Technologies}\ }\textbf {\bibinfo {volume} {5}} (\bibinfo {year}
  {2022}),\ 10.1002/qute.202100116}\BibitemShut {NoStop}%
\bibitem [{\citenamefont {Tomm}\ \emph {et~al.}(2021)\citenamefont {Tomm},
  \citenamefont {Javadi}, \citenamefont {Antoniadis}, \citenamefont {Najer},
  \citenamefont {Löbl}, \citenamefont {Korsch}, \citenamefont {Schott},
  \citenamefont {Valentin}, \citenamefont {Wieck}, \citenamefont {Ludwig},\
  and\ \citenamefont {Warburton}}]{Tomm2021}%
  \BibitemOpen
  \bibfield  {author} {\bibinfo {author} {\bibfnamefont {N.}~\bibnamefont
  {Tomm}}, \bibinfo {author} {\bibfnamefont {A.}~\bibnamefont {Javadi}},
  \bibinfo {author} {\bibfnamefont {N.~O.}\ \bibnamefont {Antoniadis}},
  \bibinfo {author} {\bibfnamefont {D.}~\bibnamefont {Najer}}, \bibinfo
  {author} {\bibfnamefont {M.~C.}\ \bibnamefont {Löbl}}, \bibinfo {author}
  {\bibfnamefont {A.~R.}\ \bibnamefont {Korsch}}, \bibinfo {author}
  {\bibfnamefont {R.}~\bibnamefont {Schott}}, \bibinfo {author} {\bibfnamefont
  {S.~R.}\ \bibnamefont {Valentin}}, \bibinfo {author} {\bibfnamefont {A.~D.}\
  \bibnamefont {Wieck}}, \bibinfo {author} {\bibfnamefont {A.}~\bibnamefont
  {Ludwig}}, \ and\ \bibinfo {author} {\bibfnamefont {R.~J.}\ \bibnamefont
  {Warburton}},\ }\href {\doibase 10.1038/s41565-020-00831-x} {\bibfield
  {journal} {\bibinfo  {journal} {Nature Nanotechnology}\ }\textbf {\bibinfo
  {volume} {16}},\ \bibinfo {pages} {399} (\bibinfo {year} {2021})}\BibitemShut
  {NoStop}%
\bibitem [{\citenamefont {Nawrath}\ \emph {et~al.}(2023)\citenamefont
  {Nawrath}, \citenamefont {Joos}, \citenamefont {Kolatschek}, \citenamefont
  {Bauer}, \citenamefont {Pruy}, \citenamefont {Hornung}, \citenamefont
  {Fischer}, \citenamefont {Huang}, \citenamefont {Vijayan}, \citenamefont
  {Sittig}, \citenamefont {Jetter}, \citenamefont {Portalupi},\ and\
  \citenamefont {Michler}}]{Nawrath2023}%
  \BibitemOpen
  \bibfield  {author} {\bibinfo {author} {\bibfnamefont {C.}~\bibnamefont
  {Nawrath}}, \bibinfo {author} {\bibfnamefont {R.}~\bibnamefont {Joos}},
  \bibinfo {author} {\bibfnamefont {S.}~\bibnamefont {Kolatschek}}, \bibinfo
  {author} {\bibfnamefont {S.}~\bibnamefont {Bauer}}, \bibinfo {author}
  {\bibfnamefont {P.}~\bibnamefont {Pruy}}, \bibinfo {author} {\bibfnamefont
  {F.}~\bibnamefont {Hornung}}, \bibinfo {author} {\bibfnamefont
  {J.}~\bibnamefont {Fischer}}, \bibinfo {author} {\bibfnamefont
  {J.}~\bibnamefont {Huang}}, \bibinfo {author} {\bibfnamefont
  {P.}~\bibnamefont {Vijayan}}, \bibinfo {author} {\bibfnamefont
  {R.}~\bibnamefont {Sittig}}, \bibinfo {author} {\bibfnamefont
  {M.}~\bibnamefont {Jetter}}, \bibinfo {author} {\bibfnamefont {S.~L.}\
  \bibnamefont {Portalupi}}, \ and\ \bibinfo {author} {\bibfnamefont
  {P.}~\bibnamefont {Michler}},\ }\href {\doibase 10.1002/qute.202300111}
  {\bibfield  {journal} {\bibinfo  {journal} {Advanced Quantum Technologies}\ }
  (\bibinfo {year} {2023}),\ 10.1002/qute.202300111}\BibitemShut {NoStop}%
\bibitem [{\citenamefont {Hanschke}\ \emph {et~al.}(2018)\citenamefont
  {Hanschke}, \citenamefont {Fischer}, \citenamefont {Appel}, \citenamefont
  {Lukin}, \citenamefont {Wierzbowski}, \citenamefont {Sun}, \citenamefont
  {Trivedi}, \citenamefont {Vučković}, \citenamefont {Finley},\ and\
  \citenamefont {Müller}}]{Hanschke2018}%
  \BibitemOpen
  \bibfield  {author} {\bibinfo {author} {\bibfnamefont {L.}~\bibnamefont
  {Hanschke}}, \bibinfo {author} {\bibfnamefont {K.~A.}\ \bibnamefont
  {Fischer}}, \bibinfo {author} {\bibfnamefont {S.}~\bibnamefont {Appel}},
  \bibinfo {author} {\bibfnamefont {D.}~\bibnamefont {Lukin}}, \bibinfo
  {author} {\bibfnamefont {J.}~\bibnamefont {Wierzbowski}}, \bibinfo {author}
  {\bibfnamefont {S.}~\bibnamefont {Sun}}, \bibinfo {author} {\bibfnamefont
  {R.}~\bibnamefont {Trivedi}}, \bibinfo {author} {\bibfnamefont
  {J.}~\bibnamefont {Vučković}}, \bibinfo {author} {\bibfnamefont {J.~J.}\
  \bibnamefont {Finley}}, \ and\ \bibinfo {author} {\bibfnamefont
  {K.}~\bibnamefont {Müller}},\ }\href {\doibase 10.1038/s41534-018-0092-0}
  {\bibfield  {journal} {\bibinfo  {journal} {npj Quantum Information}\
  }\textbf {\bibinfo {volume} {4}} (\bibinfo {year} {2018}),\
  10.1038/s41534-018-0092-0}\BibitemShut {NoStop}%
\bibitem [{\citenamefont {Santori}\ \emph {et~al.}(2002)\citenamefont
  {Santori}, \citenamefont {Fattal}, \citenamefont {Vučković}, \citenamefont
  {Solomon},\ and\ \citenamefont {Yamamoto}}]{Santori2002}%
  \BibitemOpen
  \bibfield  {author} {\bibinfo {author} {\bibfnamefont {C.}~\bibnamefont
  {Santori}}, \bibinfo {author} {\bibfnamefont {D.}~\bibnamefont {Fattal}},
  \bibinfo {author} {\bibfnamefont {J.}~\bibnamefont {Vučković}}, \bibinfo
  {author} {\bibfnamefont {G.~S.}\ \bibnamefont {Solomon}}, \ and\ \bibinfo
  {author} {\bibfnamefont {Y.}~\bibnamefont {Yamamoto}},\ }\href {\doibase 10.1038/nature01086} {\bibfield  {journal} {\bibinfo  {journal} {Nature}\
  }\textbf {\bibinfo {volume} {419}},\ \bibinfo {pages} {594} (\bibinfo {year}
  {2002})}\BibitemShut {NoStop}%
\bibitem [{\citenamefont {Somaschi}\ \emph {et~al.}(2016)\citenamefont
  {Somaschi}, \citenamefont {Giesz}, \citenamefont {Santis}, \citenamefont
  {Loredo}, \citenamefont {Almeida}, \citenamefont {Hornecker}, \citenamefont
  {Portalupi}, \citenamefont {Grange}, \citenamefont {Ant{\'{o}}n},
  \citenamefont {Demory}, \citenamefont {G{\'{o}}mez}, \citenamefont {Sagnes},
  \citenamefont {Lanzillotti-Kimura}, \citenamefont {Lema{\'{\i}}tre},
  \citenamefont {Auffeves}, \citenamefont {White}, \citenamefont {Lanco},\ and\
  \citenamefont {Senellart}}]{Somaschi2016}%
  \BibitemOpen
  \bibfield  {author} {\bibinfo {author} {\bibfnamefont {N.}~\bibnamefont
  {Somaschi}}, \bibinfo {author} {\bibfnamefont {V.}~\bibnamefont {Giesz}},
  \bibinfo {author} {\bibfnamefont {L.~D.}\ \bibnamefont {Santis}}, \bibinfo
  {author} {\bibfnamefont {J.~C.}\ \bibnamefont {Loredo}}, \bibinfo {author}
  {\bibfnamefont {M.~P.}\ \bibnamefont {Almeida}}, \bibinfo {author}
  {\bibfnamefont {G.}~\bibnamefont {Hornecker}}, \bibinfo {author}
  {\bibfnamefont {S.~L.}\ \bibnamefont {Portalupi}}, \bibinfo {author}
  {\bibfnamefont {T.}~\bibnamefont {Grange}}, \bibinfo {author} {\bibfnamefont
  {C.}~\bibnamefont {Ant{\'{o}}n}}, \bibinfo {author} {\bibfnamefont
  {J.}~\bibnamefont {Demory}}, \bibinfo {author} {\bibfnamefont
  {C.}~\bibnamefont {G{\'{o}}mez}}, \bibinfo {author} {\bibfnamefont
  {I.}~\bibnamefont {Sagnes}}, \bibinfo {author} {\bibfnamefont {N.~D.}\
  \bibnamefont {Lanzillotti-Kimura}}, \bibinfo {author} {\bibfnamefont
  {A.}~\bibnamefont {Lema{\'{\i}}tre}}, \bibinfo {author} {\bibfnamefont
  {A.}~\bibnamefont {Auffeves}}, \bibinfo {author} {\bibfnamefont {A.~G.}\
  \bibnamefont {White}}, \bibinfo {author} {\bibfnamefont {L.}~\bibnamefont
  {Lanco}}, \ and\ \bibinfo {author} {\bibfnamefont {P.}~\bibnamefont
  {Senellart}},\ }\href {\doibase 10.1038/nphoton.2016.23} {\bibfield
  {journal} {\bibinfo  {journal} {Nature Photonics}\ }\textbf {\bibinfo
  {volume} {10}},\ \bibinfo {pages} {340} (\bibinfo {year} {2016})}\BibitemShut
  {NoStop}%
\bibitem [{\citenamefont {Weiler}\ \emph {et~al.}(2010)\citenamefont {Weiler},
  \citenamefont {Ulhaq}, \citenamefont {Ulrich}, \citenamefont {Reitzenstein},
  \citenamefont {Löffler}, \citenamefont {Forchel},\ and\ \citenamefont
  {Michler}}]{Weiler2010}%
  \BibitemOpen
  \bibfield  {author} {\bibinfo {author} {\bibfnamefont {S.}~\bibnamefont
  {Weiler}}, \bibinfo {author} {\bibfnamefont {A.}~\bibnamefont {Ulhaq}},
  \bibinfo {author} {\bibfnamefont {S.~M.}\ \bibnamefont {Ulrich}}, \bibinfo
  {author} {\bibfnamefont {S.}~\bibnamefont {Reitzenstein}}, \bibinfo {author}
  {\bibfnamefont {A.}~\bibnamefont {Löffler}}, \bibinfo {author}
  {\bibfnamefont {A.}~\bibnamefont {Forchel}}, \ and\ \bibinfo {author}
  {\bibfnamefont {P.}~\bibnamefont {Michler}},\ }\href {\doibase  10.1002/pssb.201000781} {\bibfield  {journal} {\bibinfo  {journal} {physica
  status solidi (b)}\ }\textbf {\bibinfo {volume} {248}},\ \bibinfo {pages}
  {867} (\bibinfo {year} {2010})}\BibitemShut {NoStop}%
\bibitem [{\citenamefont {Purcell}(1946)}]{Purcell1946}%
  \BibitemOpen
  \bibfield  {author} {\bibinfo {author} {\bibfnamefont {E.~M.}\ \bibnamefont
  {Purcell}},\ }\href {\doibase 10.1103/physrev.69.674.2} {\bibfield  {journal}
  {\bibinfo  {journal} {Physical Review}\ }\textbf {\bibinfo {volume} {69}},\
  \bibinfo {pages} {681} (\bibinfo {year} {1946})}\BibitemShut {NoStop}%
\bibitem [{\citenamefont {Michler}(2017)}]{Michler2017}%
  \BibitemOpen
  \bibinfo {editor} {\bibfnamefont {P.}~\bibnamefont {Michler}},\ ed.,\
  \href@noop {} {\emph {\bibinfo {title} {Quantum Dots for Quantum Information
  Technologies}}},\ SpringerLink\ (\bibinfo  {publisher} {Springer},\ \bibinfo
  {address} {Cham},\ \bibinfo {year} {2017})\BibitemShut {NoStop}%
\bibitem [{\citenamefont {Ward}\ \emph {et~al.}(2005)\citenamefont {Ward},
  \citenamefont {Karimov}, \citenamefont {Unitt}, \citenamefont {Yuan},
  \citenamefont {See}, \citenamefont {Gevaux}, \citenamefont {Shields},
  \citenamefont {Atkinson},\ and\ \citenamefont {Ritchie}}]{Ward2005}%
  \BibitemOpen
  \bibfield  {author} {\bibinfo {author} {\bibfnamefont {M.~B.}\ \bibnamefont
  {Ward}}, \bibinfo {author} {\bibfnamefont {O.~Z.}\ \bibnamefont {Karimov}},
  \bibinfo {author} {\bibfnamefont {D.~C.}\ \bibnamefont {Unitt}}, \bibinfo
  {author} {\bibfnamefont {Z.~L.}\ \bibnamefont {Yuan}}, \bibinfo {author}
  {\bibfnamefont {P.}~\bibnamefont {See}}, \bibinfo {author} {\bibfnamefont
  {D.~G.}\ \bibnamefont {Gevaux}}, \bibinfo {author} {\bibfnamefont {A.~J.}\
  \bibnamefont {Shields}}, \bibinfo {author} {\bibfnamefont {P.}~\bibnamefont
  {Atkinson}}, \ and\ \bibinfo {author} {\bibfnamefont {D.~A.}\ \bibnamefont
  {Ritchie}},\ }\href {\doibase 10.1063/1.1922573} {\bibfield  {journal}
  {\bibinfo  {journal} {Applied Physics Letters}\ }\textbf {\bibinfo {volume}
  {86}} (\bibinfo {year} {2005}),\ 10.1063/1.1922573}\BibitemShut {NoStop}%
\bibitem [{\citenamefont {Paul}\ \emph {et~al.}(2015)\citenamefont {Paul},
  \citenamefont {Kettler}, \citenamefont {Zeuner}, \citenamefont {Clausen},
  \citenamefont {Jetter},\ and\ \citenamefont {Michler}}]{Paul2015}%
  \BibitemOpen
  \bibfield  {author} {\bibinfo {author} {\bibfnamefont {M.}~\bibnamefont
  {Paul}}, \bibinfo {author} {\bibfnamefont {J.}~\bibnamefont {Kettler}},
  \bibinfo {author} {\bibfnamefont {K.}~\bibnamefont {Zeuner}}, \bibinfo
  {author} {\bibfnamefont {C.}~\bibnamefont {Clausen}}, \bibinfo {author}
  {\bibfnamefont {M.}~\bibnamefont {Jetter}}, \ and\ \bibinfo {author}
  {\bibfnamefont {P.}~\bibnamefont {Michler}},\ }\href {\doibase  10.1063/1.4916349} {\bibfield  {journal} {\bibinfo  {journal} {Applied
  Physics Letters}\ }\textbf {\bibinfo {volume} {106}} (\bibinfo {year}
  {2015}),\ 10.1063/1.4916349}\BibitemShut {NoStop}%
\bibitem [{\citenamefont {Dusanowski}\ \emph {et~al.}(2017)\citenamefont
  {Dusanowski}, \citenamefont {Holewa}, \citenamefont {Maryński},
  \citenamefont {Musiał}, \citenamefont {Heuser}, \citenamefont {Srocka},
  \citenamefont {Quandt}, \citenamefont {Strittmatter}, \citenamefont {Rodt},
  \citenamefont {Misiewicz}, \citenamefont {Reitzenstein},\ and\ \citenamefont
  {S{\c e}k}}]{Dusanowski2017}%
  \BibitemOpen
  \bibfield  {author} {\bibinfo {author} {\bibfnamefont {{\L}.}~\bibnamefont
  {Dusanowski}}, \bibinfo {author} {\bibfnamefont {P.}~\bibnamefont {Holewa}},
  \bibinfo {author} {\bibfnamefont {A.}~\bibnamefont {Maryński}}, \bibinfo
  {author} {\bibfnamefont {A.}~\bibnamefont {Musiał}}, \bibinfo {author}
  {\bibfnamefont {T.}~\bibnamefont {Heuser}}, \bibinfo {author} {\bibfnamefont
  {N.}~\bibnamefont {Srocka}}, \bibinfo {author} {\bibfnamefont
  {D.}~\bibnamefont {Quandt}}, \bibinfo {author} {\bibfnamefont
  {A.}~\bibnamefont {Strittmatter}}, \bibinfo {author} {\bibfnamefont
  {S.}~\bibnamefont {Rodt}}, \bibinfo {author} {\bibfnamefont {J.}~\bibnamefont
  {Misiewicz}}, \bibinfo {author} {\bibfnamefont {S.}~\bibnamefont
  {Reitzenstein}}, \ and\ \bibinfo {author} {\bibfnamefont {G.}~\bibnamefont
  {S{\c e}k}},\ }\href {\doibase 10.1364/oe.25.031122} {\bibfield  {journal}
  {\bibinfo  {journal} {Optics Express}\ }\textbf {\bibinfo {volume} {25}},\
  \bibinfo {pages} {31122} (\bibinfo {year} {2017})}\BibitemShut {NoStop}%
\bibitem [{\citenamefont {Paul}\ \emph {et~al.}(2017)\citenamefont {Paul},
  \citenamefont {Olbrich}, \citenamefont {Höschele}, \citenamefont {Schreier},
  \citenamefont {Kettler}, \citenamefont {Portalupi}, \citenamefont {Jetter},\
  and\ \citenamefont {Michler}}]{Paul2017}%
  \BibitemOpen
  \bibfield  {author} {\bibinfo {author} {\bibfnamefont {M.}~\bibnamefont
  {Paul}}, \bibinfo {author} {\bibfnamefont {F.}~\bibnamefont {Olbrich}},
  \bibinfo {author} {\bibfnamefont {J.}~\bibnamefont {Höschele}}, \bibinfo
  {author} {\bibfnamefont {S.}~\bibnamefont {Schreier}}, \bibinfo {author}
  {\bibfnamefont {J.}~\bibnamefont {Kettler}}, \bibinfo {author} {\bibfnamefont
  {S.~L.}\ \bibnamefont {Portalupi}}, \bibinfo {author} {\bibfnamefont
  {M.}~\bibnamefont {Jetter}}, \ and\ \bibinfo {author} {\bibfnamefont
  {P.}~\bibnamefont {Michler}},\ }\href {\doibase 10.1063/1.4993935} {\bibfield
   {journal} {\bibinfo  {journal} {Applied Physics Letters}\ }\textbf {\bibinfo
  {volume} {111}} (\bibinfo {year} {2017}),\ 10.1063/1.4993935}\BibitemShut
  {NoStop}%
\bibitem [{\citenamefont {Javadi}\ \emph {et~al.}(2015)\citenamefont {Javadi},
  \citenamefont {Söllner}, \citenamefont {Arcari}, \citenamefont {Hansen},
  \citenamefont {Midolo}, \citenamefont {Mahmoodian}, \citenamefont
  {Kiršanskė}, \citenamefont {Pregnolato}, \citenamefont {Lee}, \citenamefont
  {Song}, \citenamefont {Stobbe},\ and\ \citenamefont {Lodahl}}]{Javadi2015}%
  \BibitemOpen
  \bibfield  {author} {\bibinfo {author} {\bibfnamefont {A.}~\bibnamefont
  {Javadi}}, \bibinfo {author} {\bibfnamefont {I.}~\bibnamefont {Söllner}},
  \bibinfo {author} {\bibfnamefont {M.}~\bibnamefont {Arcari}}, \bibinfo
  {author} {\bibfnamefont {S.~L.}\ \bibnamefont {Hansen}}, \bibinfo {author}
  {\bibfnamefont {L.}~\bibnamefont {Midolo}}, \bibinfo {author} {\bibfnamefont
  {S.}~\bibnamefont {Mahmoodian}}, \bibinfo {author} {\bibfnamefont
  {G.}~\bibnamefont {Kiršanskė}}, \bibinfo {author} {\bibfnamefont
  {T.}~\bibnamefont {Pregnolato}}, \bibinfo {author} {\bibfnamefont {E.~H.}\
  \bibnamefont {Lee}}, \bibinfo {author} {\bibfnamefont {J.~D.}\ \bibnamefont
  {Song}}, \bibinfo {author} {\bibfnamefont {S.}~\bibnamefont {Stobbe}}, \ and\
  \bibinfo {author} {\bibfnamefont {P.}~\bibnamefont {Lodahl}},\ }\href
  {\doibase 10.1038/ncomms9655} {\bibfield  {journal} {\bibinfo  {journal}
  {Nature Communications}\ }\textbf {\bibinfo {volume} {6}} (\bibinfo {year}
  {2015}),\ 10.1038/ncomms9655}\BibitemShut {NoStop}%
\bibitem [{\citenamefont {Liu}\ \emph {et~al.}(2019)\citenamefont {Liu},
  \citenamefont {Su}, \citenamefont {Wei}, \citenamefont {Yao}, \citenamefont
  {da~Silva}, \citenamefont {Yu}, \citenamefont {Iles-Smith}, \citenamefont
  {Srinivasan}, \citenamefont {Rastelli}, \citenamefont {Li},\ and\
  \citenamefont {Wang}}]{Liu2019}%
  \BibitemOpen
  \bibfield  {author} {\bibinfo {author} {\bibfnamefont {J.}~\bibnamefont
  {Liu}}, \bibinfo {author} {\bibfnamefont {R.}~\bibnamefont {Su}}, \bibinfo
  {author} {\bibfnamefont {Y.}~\bibnamefont {Wei}}, \bibinfo {author}
  {\bibfnamefont {B.}~\bibnamefont {Yao}}, \bibinfo {author} {\bibfnamefont
  {S.~F.~C.}\ \bibnamefont {da~Silva}}, \bibinfo {author} {\bibfnamefont
  {Y.}~\bibnamefont {Yu}}, \bibinfo {author} {\bibfnamefont {J.}~\bibnamefont
  {Iles-Smith}}, \bibinfo {author} {\bibfnamefont {K.}~\bibnamefont
  {Srinivasan}}, \bibinfo {author} {\bibfnamefont {A.}~\bibnamefont
  {Rastelli}}, \bibinfo {author} {\bibfnamefont {J.}~\bibnamefont {Li}}, \ and\
  \bibinfo {author} {\bibfnamefont {X.}~\bibnamefont {Wang}},\ }\href {\doibase 10.1038/s41565-019-0435-9} {\bibfield  {journal} {\bibinfo  {journal} {Nature
  Nanotechnology}\ }\textbf {\bibinfo {volume} {14}},\ \bibinfo {pages} {586}
  (\bibinfo {year} {2019})}\BibitemShut {NoStop}%
\bibitem [{\citenamefont {Wang}\ \emph {et~al.}(2019)\citenamefont {Wang},
  \citenamefont {Hu}, \citenamefont {Chung}, \citenamefont {Qin}, \citenamefont
  {Yang}, \citenamefont {Li}, \citenamefont {Liu}, \citenamefont {Zhong},
  \citenamefont {He}, \citenamefont {Ding}, \citenamefont {Deng}, \citenamefont
  {Dai}, \citenamefont {Huo}, \citenamefont {Höfling}, \citenamefont {Lu},\
  and\ \citenamefont {Pan}}]{Wang2019}%
  \BibitemOpen
  \bibfield  {author} {\bibinfo {author} {\bibfnamefont {H.}~\bibnamefont
  {Wang}}, \bibinfo {author} {\bibfnamefont {H.}~\bibnamefont {Hu}}, \bibinfo
  {author} {\bibfnamefont {T.-H.}\ \bibnamefont {Chung}}, \bibinfo {author}
  {\bibfnamefont {J.}~\bibnamefont {Qin}}, \bibinfo {author} {\bibfnamefont
  {X.}~\bibnamefont {Yang}}, \bibinfo {author} {\bibfnamefont {J.-P.}\
  \bibnamefont {Li}}, \bibinfo {author} {\bibfnamefont {R.-Z.}\ \bibnamefont
  {Liu}}, \bibinfo {author} {\bibfnamefont {H.-S.}\ \bibnamefont {Zhong}},
  \bibinfo {author} {\bibfnamefont {Y.-M.}\ \bibnamefont {He}}, \bibinfo
  {author} {\bibfnamefont {X.}~\bibnamefont {Ding}}, \bibinfo {author}
  {\bibfnamefont {Y.-H.}\ \bibnamefont {Deng}}, \bibinfo {author}
  {\bibfnamefont {Q.}~\bibnamefont {Dai}}, \bibinfo {author} {\bibfnamefont
  {Y.-H.}\ \bibnamefont {Huo}}, \bibinfo {author} {\bibfnamefont
  {S.}~\bibnamefont {Höfling}}, \bibinfo {author} {\bibfnamefont {C.-Y.}\
  \bibnamefont {Lu}}, \ and\ \bibinfo {author} {\bibfnamefont {J.-W.}\
  \bibnamefont {Pan}},\ }\href {\doibase 10.1103/physrevlett.122.113602}
  {\bibfield  {journal} {\bibinfo  {journal} {Physical Review Letters}\
  }\textbf {\bibinfo {volume} {122}},\ \bibinfo {pages} {113602} (\bibinfo
  {year} {2019})}\BibitemShut {NoStop}%
\bibitem [{\citenamefont {Kolatschek}\ \emph {et~al.}(2019)\citenamefont
  {Kolatschek}, \citenamefont {Hepp}, \citenamefont {Sartison}, \citenamefont
  {Jetter}, \citenamefont {Michler},\ and\ \citenamefont
  {Portalupi}}]{Kolatschek2019}%
  \BibitemOpen
  \bibfield  {author} {\bibinfo {author} {\bibfnamefont {S.}~\bibnamefont
  {Kolatschek}}, \bibinfo {author} {\bibfnamefont {S.}~\bibnamefont {Hepp}},
  \bibinfo {author} {\bibfnamefont {M.}~\bibnamefont {Sartison}}, \bibinfo
  {author} {\bibfnamefont {M.}~\bibnamefont {Jetter}}, \bibinfo {author}
  {\bibfnamefont {P.}~\bibnamefont {Michler}}, \ and\ \bibinfo {author}
  {\bibfnamefont {S.~L.}\ \bibnamefont {Portalupi}},\ }\href {\doibase  10.1063/1.5050344} {\bibfield  {journal} {\bibinfo  {journal} {Journal of
  Applied Physics}\ }\textbf {\bibinfo {volume} {125}} (\bibinfo {year}
  {2019}),\ 10.1063/1.5050344}\BibitemShut {NoStop}%
\bibitem [{\citenamefont {Phillips}\ \emph {et~al.}(2024)\citenamefont
  {Phillips}, \citenamefont {Brash}, \citenamefont {Godsland}, \citenamefont
  {Martin}, \citenamefont {Foster}, \citenamefont {Tomlinson}, \citenamefont
  {Dost}, \citenamefont {Babazadeh}, \citenamefont {Sala}, \citenamefont
  {Wilson}, \citenamefont {Heffernan}, \citenamefont {Skolnick},\ and\
  \citenamefont {Fox}}]{Phillips2024}%
  \BibitemOpen
  \bibfield  {author} {\bibinfo {author} {\bibfnamefont {C.~L.}\ \bibnamefont
  {Phillips}}, \bibinfo {author} {\bibfnamefont {A.~J.}\ \bibnamefont {Brash}},
  \bibinfo {author} {\bibfnamefont {M.}~\bibnamefont {Godsland}}, \bibinfo
  {author} {\bibfnamefont {N.~J.}\ \bibnamefont {Martin}}, \bibinfo {author}
  {\bibfnamefont {A.}~\bibnamefont {Foster}}, \bibinfo {author} {\bibfnamefont
  {A.}~\bibnamefont {Tomlinson}}, \bibinfo {author} {\bibfnamefont
  {R.}~\bibnamefont {Dost}}, \bibinfo {author} {\bibfnamefont {N.}~\bibnamefont
  {Babazadeh}}, \bibinfo {author} {\bibfnamefont {E.~M.}\ \bibnamefont {Sala}},
  \bibinfo {author} {\bibfnamefont {L.}~\bibnamefont {Wilson}}, \bibinfo
  {author} {\bibfnamefont {J.}~\bibnamefont {Heffernan}}, \bibinfo {author}
  {\bibfnamefont {M.~S.}\ \bibnamefont {Skolnick}}, \ and\ \bibinfo {author}
  {\bibfnamefont {A.~M.}\ \bibnamefont {Fox}},\ }\href {\doibase  10.1038/s41598-024-55024-6} {\bibfield  {journal} {\bibinfo  {journal}
  {Scientific Reports}\ }\textbf {\bibinfo {volume} {14}} (\bibinfo {year}
  {2024}),\ 10.1038/s41598-024-55024-6}\BibitemShut {NoStop}%
\bibitem [{\citenamefont {Herzog}\ \emph {et~al.}(2018)\citenamefont {Herzog},
  \citenamefont {Sartison}, \citenamefont {Kolatschek}, \citenamefont {Hepp},
  \citenamefont {Bommer}, \citenamefont {Pauly}, \citenamefont {Mücklich},
  \citenamefont {Becher}, \citenamefont {Jetter}, \citenamefont {Portalupi},\
  and\ \citenamefont {Michler}}]{Herzog2018}%
  \BibitemOpen
  \bibfield  {author} {\bibinfo {author} {\bibfnamefont {T.}~\bibnamefont
  {Herzog}}, \bibinfo {author} {\bibfnamefont {M.}~\bibnamefont {Sartison}},
  \bibinfo {author} {\bibfnamefont {S.}~\bibnamefont {Kolatschek}}, \bibinfo
  {author} {\bibfnamefont {S.}~\bibnamefont {Hepp}}, \bibinfo {author}
  {\bibfnamefont {A.}~\bibnamefont {Bommer}}, \bibinfo {author} {\bibfnamefont
  {C.}~\bibnamefont {Pauly}}, \bibinfo {author} {\bibfnamefont
  {F.}~\bibnamefont {Mücklich}}, \bibinfo {author} {\bibfnamefont
  {C.}~\bibnamefont {Becher}}, \bibinfo {author} {\bibfnamefont
  {M.}~\bibnamefont {Jetter}}, \bibinfo {author} {\bibfnamefont {S.~L.}\
  \bibnamefont {Portalupi}}, \ and\ \bibinfo {author} {\bibfnamefont
  {P.}~\bibnamefont {Michler}},\ }\href {\doibase 10.1088/2058-9565/aac64d}
  {\bibfield  {journal} {\bibinfo  {journal} {Quantum Science and Technology}\
  }\textbf {\bibinfo {volume} {3}},\ \bibinfo {pages} {034009} (\bibinfo {year}
  {2018})}\BibitemShut {NoStop}%
\bibitem [{\citenamefont {Pfeifer}\ \emph {et~al.}(2022)\citenamefont
  {Pfeifer}, \citenamefont {Ratschbacher}, \citenamefont {Gallego},
  \citenamefont {Saavedra}, \citenamefont {Faßbender}, \citenamefont {von
  Haaren}, \citenamefont {Alt}, \citenamefont {Hofferberth}, \citenamefont
  {Köhl}, \citenamefont {Linden},\ and\ \citenamefont
  {Meschede}}]{Pfeifer2022}%
  \BibitemOpen
  \bibfield  {author} {\bibinfo {author} {\bibfnamefont {H.}~\bibnamefont
  {Pfeifer}}, \bibinfo {author} {\bibfnamefont {L.}~\bibnamefont
  {Ratschbacher}}, \bibinfo {author} {\bibfnamefont {J.}~\bibnamefont
  {Gallego}}, \bibinfo {author} {\bibfnamefont {C.}~\bibnamefont {Saavedra}},
  \bibinfo {author} {\bibfnamefont {A.}~\bibnamefont {Faßbender}}, \bibinfo
  {author} {\bibfnamefont {A.}~\bibnamefont {von Haaren}}, \bibinfo {author}
  {\bibfnamefont {W.}~\bibnamefont {Alt}}, \bibinfo {author} {\bibfnamefont
  {S.}~\bibnamefont {Hofferberth}}, \bibinfo {author} {\bibfnamefont
  {M.}~\bibnamefont {Köhl}}, \bibinfo {author} {\bibfnamefont
  {S.}~\bibnamefont {Linden}}, \ and\ \bibinfo {author} {\bibfnamefont
  {D.}~\bibnamefont {Meschede}},\ }\href {\doibase 10.1007/s00340-022-07752-8}
  {\bibfield  {journal} {\bibinfo  {journal} {Applied Physics B}\ }\textbf
  {\bibinfo {volume} {128}} (\bibinfo {year} {2022}),\
  10.1007/s00340-022-07752-8}\BibitemShut {NoStop}%
\bibitem [{\citenamefont {Mader}\ \emph {et~al.}(2015)\citenamefont {Mader},
  \citenamefont {Reichel}, \citenamefont {H{\"a}nsch},\ and\ \citenamefont
  {Hunger}}]{Mader2015}%
  \BibitemOpen
  \bibfield  {author} {\bibinfo {author} {\bibfnamefont {M.}~\bibnamefont
  {Mader}}, \bibinfo {author} {\bibfnamefont {J.}~\bibnamefont {Reichel}},
  \bibinfo {author} {\bibfnamefont {T.~W.}\ \bibnamefont {H{\"a}nsch}}, \ and\
  \bibinfo {author} {\bibfnamefont {D.}~\bibnamefont {Hunger}},\ }\href
  {\doibase 10.1038/ncomms8249} {\bibfield  {journal} {\bibinfo  {journal}
  {Nature Communications}\ }\textbf {\bibinfo {volume} {6}},\ \bibinfo {pages}
  {7249} (\bibinfo {year} {2015})}\BibitemShut {NoStop}%
\bibitem [{\citenamefont {Schlehahn}\ \emph {et~al.}(2018)\citenamefont
  {Schlehahn}, \citenamefont {Fischbach}, \citenamefont {Schmidt},
  \citenamefont {Kaganskiy}, \citenamefont {Strittmatter}, \citenamefont
  {Rodt}, \citenamefont {Heindel},\ and\ \citenamefont
  {Reitzenstein}}]{Schlehahn2018}%
  \BibitemOpen
  \bibfield  {author} {\bibinfo {author} {\bibfnamefont {A.}~\bibnamefont
  {Schlehahn}}, \bibinfo {author} {\bibfnamefont {S.}~\bibnamefont
  {Fischbach}}, \bibinfo {author} {\bibfnamefont {R.}~\bibnamefont {Schmidt}},
  \bibinfo {author} {\bibfnamefont {A.}~\bibnamefont {Kaganskiy}}, \bibinfo
  {author} {\bibfnamefont {A.}~\bibnamefont {Strittmatter}}, \bibinfo {author}
  {\bibfnamefont {S.}~\bibnamefont {Rodt}}, \bibinfo {author} {\bibfnamefont
  {T.}~\bibnamefont {Heindel}}, \ and\ \bibinfo {author} {\bibfnamefont
  {S.}~\bibnamefont {Reitzenstein}},\ }\href {\doibase  10.1038/s41598-017-19049-4} {\bibfield  {journal} {\bibinfo  {journal}
  {Scientific Reports}\ }\textbf {\bibinfo {volume} {8}},\ \bibinfo {pages}
  {1340} (\bibinfo {year} {2018})}\BibitemShut {NoStop}%
\bibitem [{\citenamefont {Rickert}\ \emph {et~al.}(2019)\citenamefont
  {Rickert}, \citenamefont {Kupko}, \citenamefont {Rodt}, \citenamefont
  {Reitzenstein},\ and\ \citenamefont {Heindel}}]{Rickert19}%
  \BibitemOpen
  \bibfield  {author} {\bibinfo {author} {\bibfnamefont {L.}~\bibnamefont
  {Rickert}}, \bibinfo {author} {\bibfnamefont {T.}~\bibnamefont {Kupko}},
  \bibinfo {author} {\bibfnamefont {S.}~\bibnamefont {Rodt}}, \bibinfo {author}
  {\bibfnamefont {S.}~\bibnamefont {Reitzenstein}}, \ and\ \bibinfo {author}
  {\bibfnamefont {T.}~\bibnamefont {Heindel}},\ }\href {\doibase  10.1364/OE.27.036824} {\bibfield  {journal} {\bibinfo  {journal} {Opt.
  Express}\ }\textbf {\bibinfo {volume} {27}},\ \bibinfo {pages} {36824}
  (\bibinfo {year} {2019})}\BibitemShut {NoStop}%
\bibitem [{\citenamefont {Rickert}\ \emph {et~al.}(2021)\citenamefont
  {Rickert}, \citenamefont {Schröder}, \citenamefont {Gao}, \citenamefont
  {Schneider}, \citenamefont {Höfling},\ and\ \citenamefont
  {Heindel}}]{Rickert21}%
  \BibitemOpen
  \bibfield  {author} {\bibinfo {author} {\bibfnamefont {L.}~\bibnamefont
  {Rickert}}, \bibinfo {author} {\bibfnamefont {F.}~\bibnamefont {Schröder}},
  \bibinfo {author} {\bibfnamefont {T.}~\bibnamefont {Gao}}, \bibinfo {author}
  {\bibfnamefont {C.}~\bibnamefont {Schneider}}, \bibinfo {author}
  {\bibfnamefont {S.}~\bibnamefont {Höfling}}, \ and\ \bibinfo {author}
  {\bibfnamefont {T.}~\bibnamefont {Heindel}},\ }\href {\doibase  10.1063/5.0063697} {\bibfield  {journal} {\bibinfo  {journal} {Applied
  Physics Letters}\ }\textbf {\bibinfo {volume} {119}},\ \bibinfo {pages}
  {131104} (\bibinfo {year} {2021})},\ \Eprint
  {http://arxiv.org/abs/https://pubs.aip.org/aip/apl/article-pdf/doi/10.1063/5.0063697/14553180/131104\_1\_online.pdf}{https://pubs.aip.org/aip/apl/article-pdf/doi/10.1063/5.0063697/14553180/131104\_1\_online.pdf}\BibitemShut
  {NoStop}%
\bibitem [{\citenamefont {Ding}\ \emph {et~al.}(2023)\citenamefont {Ding},
  \citenamefont {Guo}, \citenamefont {Xu}, \citenamefont {Liu}, \citenamefont
  {Zou}, \citenamefont {Zhao}, \citenamefont {Ge}, \citenamefont {Zhang},
  \citenamefont {Liu}, \citenamefont {Wang}, \citenamefont {Chen},
  \citenamefont {Wang}, \citenamefont {He}, \citenamefont {Huo}, \citenamefont
  {Lu},\ and\ \citenamefont {Pan}}]{Ding2023}%
  \BibitemOpen
  \bibfield  {author} {\bibinfo {author} {\bibfnamefont {X.}~\bibnamefont
  {Ding}}, \bibinfo {author} {\bibfnamefont {Y.-P.}\ \bibnamefont {Guo}},
  \bibinfo {author} {\bibfnamefont {M.-C.}\ \bibnamefont {Xu}}, \bibinfo
  {author} {\bibfnamefont {R.-Z.}\ \bibnamefont {Liu}}, \bibinfo {author}
  {\bibfnamefont {G.-Y.}\ \bibnamefont {Zou}}, \bibinfo {author} {\bibfnamefont
  {J.-Y.}\ \bibnamefont {Zhao}}, \bibinfo {author} {\bibfnamefont {Z.-X.}\
  \bibnamefont {Ge}}, \bibinfo {author} {\bibfnamefont {Q.-H.}\ \bibnamefont
  {Zhang}}, \bibinfo {author} {\bibfnamefont {H.-L.}\ \bibnamefont {Liu}},
  \bibinfo {author} {\bibfnamefont {L.-J.}\ \bibnamefont {Wang}}, \bibinfo
  {author} {\bibfnamefont {M.-C.}\ \bibnamefont {Chen}}, \bibinfo {author}
  {\bibfnamefont {H.}~\bibnamefont {Wang}}, \bibinfo {author} {\bibfnamefont
  {Y.-M.}\ \bibnamefont {He}}, \bibinfo {author} {\bibfnamefont {Y.-H.}\
  \bibnamefont {Huo}}, \bibinfo {author} {\bibfnamefont {C.-Y.}\ \bibnamefont
  {Lu}}, \ and\ \bibinfo {author} {\bibfnamefont {J.-W.}\ \bibnamefont {Pan}},\
  }\href {\doibase 10.48550/ARXIV.2311.08347} {\  (\bibinfo {year} {2023}),\
  10.48550/ARXIV.2311.08347},\ \Eprint
  {http://arxiv.org/abs/2311.08347}{arXiv:2311.08347 [quant-ph]}\BibitemShut
  {NoStop}%
\bibitem [{\citenamefont {Kolatschek}\ \emph {et~al.}(2021)\citenamefont
  {Kolatschek}, \citenamefont {Nawrath}, \citenamefont {Bauer}, \citenamefont
  {Huang}, \citenamefont {Fischer}, \citenamefont {Sittig}, \citenamefont
  {Jetter}, \citenamefont {Portalupi},\ and\ \citenamefont
  {Michler}}]{Kolatschek2021}%
  \BibitemOpen
  \bibfield  {author} {\bibinfo {author} {\bibfnamefont {S.}~\bibnamefont
  {Kolatschek}}, \bibinfo {author} {\bibfnamefont {C.}~\bibnamefont {Nawrath}},
  \bibinfo {author} {\bibfnamefont {S.}~\bibnamefont {Bauer}}, \bibinfo
  {author} {\bibfnamefont {J.}~\bibnamefont {Huang}}, \bibinfo {author}
  {\bibfnamefont {J.}~\bibnamefont {Fischer}}, \bibinfo {author} {\bibfnamefont
  {R.}~\bibnamefont {Sittig}}, \bibinfo {author} {\bibfnamefont
  {M.}~\bibnamefont {Jetter}}, \bibinfo {author} {\bibfnamefont {S.~L.}\
  \bibnamefont {Portalupi}}, \ and\ \bibinfo {author} {\bibfnamefont
  {P.}~\bibnamefont {Michler}},\ }\href {\doibase 10.1021/acs.nanolett.1c02647}
  {\bibfield  {journal} {\bibinfo  {journal} {Nano Letters}\ }\textbf {\bibinfo
  {volume} {21}},\ \bibinfo {pages} {7740} (\bibinfo {year}
  {2021})}\BibitemShut {NoStop}%
\bibitem [{\citenamefont {Hunger}\ \emph {et~al.}(2010)\citenamefont {Hunger},
  \citenamefont {Steinmetz}, \citenamefont {Colombe}, \citenamefont {Deutsch},
  \citenamefont {Hänsch},\ and\ \citenamefont {Reichel}}]{Hunger2010}%
  \BibitemOpen
  \bibfield  {author} {\bibinfo {author} {\bibfnamefont {D.}~\bibnamefont
  {Hunger}}, \bibinfo {author} {\bibfnamefont {T.}~\bibnamefont {Steinmetz}},
  \bibinfo {author} {\bibfnamefont {Y.}~\bibnamefont {Colombe}}, \bibinfo
  {author} {\bibfnamefont {C.}~\bibnamefont {Deutsch}}, \bibinfo {author}
  {\bibfnamefont {T.~W.}\ \bibnamefont {Hänsch}}, \ and\ \bibinfo {author}
  {\bibfnamefont {J.}~\bibnamefont {Reichel}},\ }\href {\doibase  10.1088/1367-2630/12/6/065038} {\bibfield  {journal} {\bibinfo  {journal}
  {New Journal of Physics}\ }\textbf {\bibinfo {volume} {12}},\ \bibinfo
  {pages} {065038} (\bibinfo {year} {2010})}\BibitemShut {NoStop}%
\bibitem [{\citenamefont {Janitz}\ \emph {et~al.}(2015)\citenamefont {Janitz},
  \citenamefont {Ruf}, \citenamefont {Dimock}, \citenamefont {Bourassa},
  \citenamefont {Sankey},\ and\ \citenamefont {Childress}}]{Janitz2015}%
  \BibitemOpen
  \bibfield  {author} {\bibinfo {author} {\bibfnamefont {E.}~\bibnamefont
  {Janitz}}, \bibinfo {author} {\bibfnamefont {M.}~\bibnamefont {Ruf}},
  \bibinfo {author} {\bibfnamefont {M.}~\bibnamefont {Dimock}}, \bibinfo
  {author} {\bibfnamefont {A.}~\bibnamefont {Bourassa}}, \bibinfo {author}
  {\bibfnamefont {J.}~\bibnamefont {Sankey}}, \ and\ \bibinfo {author}
  {\bibfnamefont {L.}~\bibnamefont {Childress}},\ }\href {\doibase  10.1103/physreva.92.043844} {\bibfield  {journal} {\bibinfo  {journal}
  {Physical Review A}\ }\textbf {\bibinfo {volume} {92}},\ \bibinfo {pages}
  {043844} (\bibinfo {year} {2015})}\BibitemShut {NoStop}%
\bibitem [{\citenamefont {van Dam}\ \emph {et~al.}(2018)\citenamefont {van
  Dam}, \citenamefont {Ruf},\ and\ \citenamefont {Hanson}}]{Dam2018}%
  \BibitemOpen
  \bibfield  {author} {\bibinfo {author} {\bibfnamefont {S.~B.}\ \bibnamefont
  {van Dam}}, \bibinfo {author} {\bibfnamefont {M.}~\bibnamefont {Ruf}}, \ and\
  \bibinfo {author} {\bibfnamefont {R.}~\bibnamefont {Hanson}},\ }\href
  {\doibase 10.1088/1367-2630/aaec29} {\bibfield  {journal} {\bibinfo
  {journal} {New Journal of Physics}\ }\textbf {\bibinfo {volume} {20}},\
  \bibinfo {pages} {115004} (\bibinfo {year} {2018})}\BibitemShut {NoStop}%
\bibitem [{\citenamefont {Sartison}\ \emph {et~al.}(2017)\citenamefont
  {Sartison}, \citenamefont {Portalupi}, \citenamefont {Gissibl}, \citenamefont
  {Jetter}, \citenamefont {Giessen},\ and\ \citenamefont
  {Michler}}]{Sartison2017}%
  \BibitemOpen
  \bibfield  {author} {\bibinfo {author} {\bibfnamefont {M.}~\bibnamefont
  {Sartison}}, \bibinfo {author} {\bibfnamefont {S.~L.}\ \bibnamefont
  {Portalupi}}, \bibinfo {author} {\bibfnamefont {T.}~\bibnamefont {Gissibl}},
  \bibinfo {author} {\bibfnamefont {M.}~\bibnamefont {Jetter}}, \bibinfo
  {author} {\bibfnamefont {H.}~\bibnamefont {Giessen}}, \ and\ \bibinfo
  {author} {\bibfnamefont {P.}~\bibnamefont {Michler}},\ }\href {\doibase  10.1038/srep39916} {\bibfield  {journal} {\bibinfo  {journal} {Scientific
  Reports}\ }\textbf {\bibinfo {volume} {7}} (\bibinfo {year} {2017}),\
  10.1038/srep39916}\BibitemShut {NoStop}%
\bibitem [{\citenamefont {Greuter}\ \emph {et~al.}(2014)\citenamefont
  {Greuter}, \citenamefont {Starosielec}, \citenamefont {Najer}, \citenamefont
  {Ludwig}, \citenamefont {Duempelmann}, \citenamefont {Rohner},\ and\
  \citenamefont {Warburton}}]{Greuter2014}%
  \BibitemOpen
  \bibfield  {author} {\bibinfo {author} {\bibfnamefont {L.}~\bibnamefont
  {Greuter}}, \bibinfo {author} {\bibfnamefont {S.}~\bibnamefont
  {Starosielec}}, \bibinfo {author} {\bibfnamefont {D.}~\bibnamefont {Najer}},
  \bibinfo {author} {\bibfnamefont {A.}~\bibnamefont {Ludwig}}, \bibinfo
  {author} {\bibfnamefont {L.}~\bibnamefont {Duempelmann}}, \bibinfo {author}
  {\bibfnamefont {D.}~\bibnamefont {Rohner}}, \ and\ \bibinfo {author}
  {\bibfnamefont {R.~J.}\ \bibnamefont {Warburton}},\ }\href {\doibase  10.1063/1.4896415} {\bibfield  {journal} {\bibinfo  {journal} {Applied
  Physics Letters}\ }\textbf {\bibinfo {volume} {105}} (\bibinfo {year}
  {2014}),\ 10.1063/1.4896415}\BibitemShut {NoStop}%
\bibitem [{\citenamefont {Koks}\ and\ \citenamefont {van
  Exter}(2021)}]{Koks2021}%
  \BibitemOpen
  \bibfield  {author} {\bibinfo {author} {\bibfnamefont {C.}~\bibnamefont
  {Koks}}\ and\ \bibinfo {author} {\bibfnamefont {M.~P.}\ \bibnamefont {van
  Exter}},\ }\href {\doibase 10.1364/oe.412346} {\bibfield  {journal} {\bibinfo
   {journal} {Optics Express}\ }\textbf {\bibinfo {volume} {29}},\ \bibinfo
  {pages} {6879} (\bibinfo {year} {2021})}\BibitemShut {NoStop}%
\bibitem [{\citenamefont {Fontana}\ \emph {et~al.}(2021)\citenamefont
  {Fontana}, \citenamefont {Zifkin}, \citenamefont {Janitz}, \citenamefont
  {Rodríguez~Rosenblueth},\ and\ \citenamefont {Childress}}]{Fontana2021}%
  \BibitemOpen
  \bibfield  {author} {\bibinfo {author} {\bibfnamefont {Y.}~\bibnamefont
  {Fontana}}, \bibinfo {author} {\bibfnamefont {R.}~\bibnamefont {Zifkin}},
  \bibinfo {author} {\bibfnamefont {E.}~\bibnamefont {Janitz}}, \bibinfo
  {author} {\bibfnamefont {C.~D.}\ \bibnamefont {Rodríguez~Rosenblueth}}, \
  and\ \bibinfo {author} {\bibfnamefont {L.}~\bibnamefont {Childress}},\ }\href
  {\doibase 10.1063/5.0049520} {\bibfield  {journal} {\bibinfo  {journal}
  {Review of Scientific Instruments}\ }\textbf {\bibinfo {volume} {92}}
  (\bibinfo {year} {2021}),\ 10.1063/5.0049520}\BibitemShut {NoStop}%
\bibitem [{\citenamefont {Pallmann}\ \emph {et~al.}(2023)\citenamefont
  {Pallmann}, \citenamefont {Eichhorn}, \citenamefont {Benedikter},
  \citenamefont {Casabone}, \citenamefont {Hümmer},\ and\ \citenamefont
  {Hunger}}]{Pallmann2023}%
  \BibitemOpen
  \bibfield  {author} {\bibinfo {author} {\bibfnamefont {M.}~\bibnamefont
  {Pallmann}}, \bibinfo {author} {\bibfnamefont {T.}~\bibnamefont {Eichhorn}},
  \bibinfo {author} {\bibfnamefont {J.}~\bibnamefont {Benedikter}}, \bibinfo
  {author} {\bibfnamefont {B.}~\bibnamefont {Casabone}}, \bibinfo {author}
  {\bibfnamefont {T.}~\bibnamefont {Hümmer}}, \ and\ \bibinfo {author}
  {\bibfnamefont {D.}~\bibnamefont {Hunger}},\ }\href {\doibase 10.1063/5.0139003} {\bibfield  {journal} {\bibinfo  {journal} {{APL}
  Photonics}\ }\textbf {\bibinfo {volume} {8}} (\bibinfo {year} {2023}),\
  10.1063/5.0139003}\BibitemShut {NoStop}%
\bibitem [{\citenamefont {Fisicaro}\ \emph {et~al.}(2024)\citenamefont
  {Fisicaro}, \citenamefont {Witlox}, \citenamefont {van~der Meer},\ and\
  \citenamefont {Löffler}}]{Fisicaro2024}%
  \BibitemOpen
  \bibfield  {author} {\bibinfo {author} {\bibfnamefont {M.}~\bibnamefont
  {Fisicaro}}, \bibinfo {author} {\bibfnamefont {M.}~\bibnamefont {Witlox}},
  \bibinfo {author} {\bibfnamefont {H.}~\bibnamefont {van~der Meer}}, \ and\
  \bibinfo {author} {\bibfnamefont {W.}~\bibnamefont {Löffler}},\ }\href
  {\doibase 10.1063/5.0174982} {\bibfield  {journal} {\bibinfo  {journal}
  {Review of Scientific Instruments}\ }\textbf {\bibinfo {volume} {95}}
  (\bibinfo {year} {2024}),\ 10.1063/5.0174982}\BibitemShut {NoStop}%
\bibitem [{\citenamefont {Gallego}\ \emph {et~al.}(2016)\citenamefont
  {Gallego}, \citenamefont {Ghosh}, \citenamefont {Alavi}, \citenamefont {Alt},
  \citenamefont {Martinez-Dorantes}, \citenamefont {Meschede},\ and\
  \citenamefont {Ratschbacher}}]{Gallego2016}%
  \BibitemOpen
  \bibfield  {author} {\bibinfo {author} {\bibfnamefont {J.}~\bibnamefont
  {Gallego}}, \bibinfo {author} {\bibfnamefont {S.}~\bibnamefont {Ghosh}},
  \bibinfo {author} {\bibfnamefont {S.~K.}\ \bibnamefont {Alavi}}, \bibinfo
  {author} {\bibfnamefont {W.}~\bibnamefont {Alt}}, \bibinfo {author}
  {\bibfnamefont {M.}~\bibnamefont {Martinez-Dorantes}}, \bibinfo {author}
  {\bibfnamefont {D.}~\bibnamefont {Meschede}}, \ and\ \bibinfo {author}
  {\bibfnamefont {L.}~\bibnamefont {Ratschbacher}},\ }\href {\doibase  10.1007/s00340-015-6281-z} {\bibfield  {journal} {\bibinfo  {journal}
  {Applied Physics B}\ }\textbf {\bibinfo {volume} {122}} (\bibinfo {year}
  {2016}),\ 10.1007/s00340-015-6281-z}\BibitemShut {NoStop}%
\bibitem [{\citenamefont {Saleh}\ and\ \citenamefont
  {Teich}(1991)}]{Saleh1991}%
  \BibitemOpen
  \bibfield  {author} {\bibinfo {author} {\bibfnamefont {B.~E.~A.}\
  \bibnamefont {Saleh}}\ and\ \bibinfo {author} {\bibfnamefont {M.~C.}\
  \bibnamefont {Teich}},\ }\href {\doibase 10.1002/0471213748} {\emph {\bibinfo
  {title} {Fundamentals of Photonics}}}\ (\bibinfo  {publisher} {Wiley},\
  \bibinfo {year} {1991})\BibitemShut {NoStop}%
\bibitem [{\citenamefont {Bennett}\ and\ \citenamefont
  {Porteus}(1961)}]{Bennett1961}%
  \BibitemOpen
  \bibfield  {author} {\bibinfo {author} {\bibfnamefont {H.~E.}\ \bibnamefont
  {Bennett}}\ and\ \bibinfo {author} {\bibfnamefont {J.~O.}\ \bibnamefont
  {Porteus}},\ }\href {\doibase 10.1364/josa.51.000123} {\bibfield  {journal}
  {\bibinfo  {journal} {Journal of the Optical Society of America}\ }\textbf
  {\bibinfo {volume} {51}},\ \bibinfo {pages} {123} (\bibinfo {year}
  {1961})}\BibitemShut {NoStop}%
\bibitem [{\citenamefont {Grange}\ \emph {et~al.}(2015)\citenamefont {Grange},
  \citenamefont {Hornecker}, \citenamefont {Hunger}, \citenamefont {Poizat},
  \citenamefont {Gérard}, \citenamefont {Senellart},\ and\ \citenamefont
  {Auffèves}}]{Grange2015}%
  \BibitemOpen
  \bibfield  {author} {\bibinfo {author} {\bibfnamefont {T.}~\bibnamefont
  {Grange}}, \bibinfo {author} {\bibfnamefont {G.}~\bibnamefont {Hornecker}},
  \bibinfo {author} {\bibfnamefont {D.}~\bibnamefont {Hunger}}, \bibinfo
  {author} {\bibfnamefont {J.-P.}\ \bibnamefont {Poizat}}, \bibinfo {author}
  {\bibfnamefont {J.-M.}\ \bibnamefont {Gérard}}, \bibinfo {author}
  {\bibfnamefont {P.}~\bibnamefont {Senellart}}, \ and\ \bibinfo {author}
  {\bibfnamefont {A.}~\bibnamefont {Auffèves}},\ }\href {\doibase  10.1103/physrevlett.114.193601} {\bibfield  {journal} {\bibinfo  {journal}
  {Physical Review Letters}\ }\textbf {\bibinfo {volume} {114}},\ \bibinfo
  {pages} {193601} (\bibinfo {year} {2015})}\BibitemShut {NoStop}%
\bibitem [{\citenamefont {Hood}\ \emph {et~al.}(2001)\citenamefont {Hood},
  \citenamefont {Kimble},\ and\ \citenamefont {Ye}}]{Hood2001}%
  \BibitemOpen
  \bibfield  {author} {\bibinfo {author} {\bibfnamefont {C.~J.}\ \bibnamefont
  {Hood}}, \bibinfo {author} {\bibfnamefont {H.~J.}\ \bibnamefont {Kimble}}, \
  and\ \bibinfo {author} {\bibfnamefont {J.}~\bibnamefont {Ye}},\ }\href
  {\doibase 10.1103/physreva.64.033804} {\bibfield  {journal} {\bibinfo
  {journal} {Physical Review A}\ }\textbf {\bibinfo {volume} {64}},\ \bibinfo
  {pages} {033804} (\bibinfo {year} {2001})}\BibitemShut {NoStop}%
\bibitem [{\citenamefont {Novotny}(2012)}]{Novotny2012}%
  \BibitemOpen
  \bibfield  {author} {\bibinfo {author} {\bibfnamefont {L.}~\bibnamefont
  {Novotny}},\ }\href@noop {} {\emph {\bibinfo {title} {Principles of
  nano-optics}}},\ \bibinfo {edition} {2nd}\ ed.,\ edited by\ \bibinfo {editor}
  {\bibfnamefont {B.}~\bibnamefont {Hecht}}\ (\bibinfo  {publisher} {Cambridge
  University Press},\ \bibinfo {address} {Cambridge},\ \bibinfo {year} {2012})\
  \bibinfo {note} {description based on print version record}\BibitemShut
  {NoStop}%
\bibitem [{\citenamefont {Auffèves}\ \emph {et~al.}(2010)\citenamefont
  {Auffèves}, \citenamefont {Gerace}, \citenamefont {Gérard}, \citenamefont
  {Santos}, \citenamefont {Andreani},\ and\ \citenamefont
  {Poizat}}]{Auffeves2010}%
  \BibitemOpen
  \bibfield  {author} {\bibinfo {author} {\bibfnamefont {A.}~\bibnamefont
  {Auffèves}}, \bibinfo {author} {\bibfnamefont {D.}~\bibnamefont {Gerace}},
  \bibinfo {author} {\bibfnamefont {J.-M.}\ \bibnamefont {Gérard}}, \bibinfo
  {author} {\bibfnamefont {M.~F.}\ \bibnamefont {Santos}}, \bibinfo {author}
  {\bibfnamefont {L.~C.}\ \bibnamefont {Andreani}}, \ and\ \bibinfo {author}
  {\bibfnamefont {J.-P.}\ \bibnamefont {Poizat}},\ }\href {\doibase  10.1103/physrevb.81.245419} {\bibfield  {journal} {\bibinfo  {journal}
  {Physical Review B}\ }\textbf {\bibinfo {volume} {81}},\ \bibinfo {pages}
  {245419} (\bibinfo {year} {2010})}\BibitemShut {NoStop}%
\bibitem [{\citenamefont {Zifkin}\ \emph {et~al.}(2023)\citenamefont {Zifkin},
  \citenamefont {Rosenblueth}, \citenamefont {Janitz}, \citenamefont
  {Fontana},\ and\ \citenamefont {Childress}}]{Zifkin2023}%
  \BibitemOpen
  \bibfield  {author} {\bibinfo {author} {\bibfnamefont {R.}~\bibnamefont
  {Zifkin}}, \bibinfo {author} {\bibfnamefont {C.~D.~R.}\ \bibnamefont
  {Rosenblueth}}, \bibinfo {author} {\bibfnamefont {E.}~\bibnamefont {Janitz}},
  \bibinfo {author} {\bibfnamefont {Y.}~\bibnamefont {Fontana}}, \ and\
  \bibinfo {author} {\bibfnamefont {L.}~\bibnamefont {Childress}},\ }\href
  {\doibase 10.48550/ARXIV.2312.14313} {\  (\bibinfo {year} {2023}),\
  10.48550/ARXIV.2312.14313},\ \Eprint
  {http://arxiv.org/abs/2312.14313}{arXiv:2312.14313 [quant-ph]}\BibitemShut
  {NoStop}%
\bibitem [{\citenamefont {Guha}\ \emph {et~al.}(2017)\citenamefont {Guha},
  \citenamefont {Marsault}, \citenamefont {Cadiz}, \citenamefont {Morgenroth},
  \citenamefont {Ulin}, \citenamefont {Berkovitz}, \citenamefont {Lemaître},
  \citenamefont {Gomez}, \citenamefont {Amo}, \citenamefont {Combrié},
  \citenamefont {Gérard}, \citenamefont {Leo},\ and\ \citenamefont
  {Favero}}]{Guha2017}%
  \BibitemOpen
  \bibfield  {author} {\bibinfo {author} {\bibfnamefont {B.}~\bibnamefont
  {Guha}}, \bibinfo {author} {\bibfnamefont {F.}~\bibnamefont {Marsault}},
  \bibinfo {author} {\bibfnamefont {F.}~\bibnamefont {Cadiz}}, \bibinfo
  {author} {\bibfnamefont {L.}~\bibnamefont {Morgenroth}}, \bibinfo {author}
  {\bibfnamefont {V.}~\bibnamefont {Ulin}}, \bibinfo {author} {\bibfnamefont
  {V.}~\bibnamefont {Berkovitz}}, \bibinfo {author} {\bibfnamefont
  {A.}~\bibnamefont {Lemaître}}, \bibinfo {author} {\bibfnamefont
  {C.}~\bibnamefont {Gomez}}, \bibinfo {author} {\bibfnamefont
  {A.}~\bibnamefont {Amo}}, \bibinfo {author} {\bibfnamefont {S.}~\bibnamefont
  {Combrié}}, \bibinfo {author} {\bibfnamefont {B.}~\bibnamefont {Gérard}},
  \bibinfo {author} {\bibfnamefont {G.}~\bibnamefont {Leo}}, \ and\ \bibinfo
  {author} {\bibfnamefont {I.}~\bibnamefont {Favero}},\ }\href {\doibase  10.1364/optica.4.000218} {\bibfield  {journal} {\bibinfo  {journal} {Optica}\
  }\textbf {\bibinfo {volume} {4}},\ \bibinfo {pages} {218} (\bibinfo {year}
  {2017})}\BibitemShut {NoStop}%
\bibitem [{\citenamefont {Najer}\ \emph {et~al.}(2019)\citenamefont {Najer},
  \citenamefont {Söllner}, \citenamefont {Sekatski}, \citenamefont {Dolique},
  \citenamefont {Löbl}, \citenamefont {Riedel}, \citenamefont {Schott},
  \citenamefont {Starosielec}, \citenamefont {Valentin}, \citenamefont {Wieck},
  \citenamefont {Sangouard}, \citenamefont {Ludwig},\ and\ \citenamefont
  {Warburton}}]{Najer2019}%
  \BibitemOpen
  \bibfield  {author} {\bibinfo {author} {\bibfnamefont {D.}~\bibnamefont
  {Najer}}, \bibinfo {author} {\bibfnamefont {I.}~\bibnamefont {Söllner}},
  \bibinfo {author} {\bibfnamefont {P.}~\bibnamefont {Sekatski}}, \bibinfo
  {author} {\bibfnamefont {V.}~\bibnamefont {Dolique}}, \bibinfo {author}
  {\bibfnamefont {M.~C.}\ \bibnamefont {Löbl}}, \bibinfo {author}
  {\bibfnamefont {D.}~\bibnamefont {Riedel}}, \bibinfo {author} {\bibfnamefont
  {R.}~\bibnamefont {Schott}}, \bibinfo {author} {\bibfnamefont
  {S.}~\bibnamefont {Starosielec}}, \bibinfo {author} {\bibfnamefont {S.~R.}\
  \bibnamefont {Valentin}}, \bibinfo {author} {\bibfnamefont {A.~D.}\
  \bibnamefont {Wieck}}, \bibinfo {author} {\bibfnamefont {N.}~\bibnamefont
  {Sangouard}}, \bibinfo {author} {\bibfnamefont {A.}~\bibnamefont {Ludwig}}, \
  and\ \bibinfo {author} {\bibfnamefont {R.~J.}\ \bibnamefont {Warburton}},\
  }\href {\doibase 10.1038/s41586-019-1709-y} {\bibfield  {journal} {\bibinfo
  {journal} {Nature}\ }\textbf {\bibinfo {volume} {575}},\ \bibinfo {pages}
  {622} (\bibinfo {year} {2019})}\BibitemShut {NoStop}%
\bibitem [{\citenamefont {Fox}(2006)}]{Fox2006}%
  \BibitemOpen
  \bibfield  {author} {\bibinfo {author} {\bibfnamefont {M.}~\bibnamefont
  {Fox}},\ }\href@noop {} {\emph {\bibinfo {title} {Quantum optics}}}\
  (\bibinfo  {publisher} {Oxford University Press},\ \bibinfo {year} {2006})\
  p.\ \bibinfo {pages} {378}\BibitemShut {NoStop}%
\bibitem [{\citenamefont {Furtak-Wells}\ \emph {et~al.}(2018)\citenamefont
  {Furtak-Wells}, \citenamefont {Clark}, \citenamefont {Purdy},\ and\
  \citenamefont {Beige}}]{FurtakWells2018}%
  \BibitemOpen
  \bibfield  {author} {\bibinfo {author} {\bibfnamefont {N.}~\bibnamefont
  {Furtak-Wells}}, \bibinfo {author} {\bibfnamefont {L.~A.}\ \bibnamefont
  {Clark}}, \bibinfo {author} {\bibfnamefont {R.}~\bibnamefont {Purdy}}, \ and\
  \bibinfo {author} {\bibfnamefont {A.}~\bibnamefont {Beige}},\ }\href
  {\doibase 10.1103/physreva.97.043827} {\bibfield  {journal} {\bibinfo
  {journal} {Physical Review A}\ }\textbf {\bibinfo {volume} {97}},\ \bibinfo
  {pages} {043827} (\bibinfo {year} {2018})}\BibitemShut {NoStop}%
\end{thebibliography}

%

\end{document}